\documentclass[graybox]{svmult}

\usepackage{mathptmx}       
\usepackage{helvet}         
\usepackage{courier}        
\usepackage{type1cm}        
\usepackage{makeidx}         
\usepackage{graphicx}        
\usepackage{multicol}        
\usepackage[bottom]{footmisc}

\usepackage{amssymb}
\usepackage{amsmath}

\makeindex             

\newcommand{\avg}[1]{\ensuremath{\langle #1 \rangle}}
\newcommand{\bma}{\begin{math}}
\newcommand{\ema}{\end{math}}
\newcommand{\beq}{\begin{equation}}
\newcommand{\eeq}{\end{equation}}
\newcommand{\beqa}{\begin{eqnarray}}
\newcommand{\eeqa}{\end{eqnarray}}
\newcommand{\bc}{\begin{center}}
\newcommand{\ec}{\end{center}} 
\newcommand{\bit}{\begin{itemize}}
\newcommand{\eit}{\end{itemize}}

\begin{document}

\title*{Modeling the Intergalactic Medium during the Epoch of Reionization}
\author{Adam Lidz}
\institute{Adam Lidz \at University of Pennsylvania, Department of Physics \& Astronomy, 209 S. 33rd Street, Philadelphia, PA 19104, \email{alidz@sas.upenn.edu}}
%
%
\maketitle


\abstract{A major goal of observational and theoretical cosmology is to observe the largely unexplored time period in the history of our universe when the first
galaxies form, and to interpret these measurements. Early galaxies dramatically impacted the gas around them in the surrounding intergalactic medium (IGM) by
photoionzing the gas during the ``Epoch of Reionization'' (EoR). This epoch likely spanned an extended stretch in cosmic time: ionized regions formed and grew around early generations
of galaxies, gradually filling a larger and larger fraction of the volume of the universe. At some time -- thus far uncertain, but within the first billion years or so after the big bang -- essentially
the entire volume of the universe became filled with ionized gas. 
The properties of the IGM provide valuable information regarding the formation time and nature of early galaxy populations, and many approaches for studying the first luminous sources are hence based on measurements of the surrounding intergalactic gas. 
The prospects for improved reionization-era observations of the IGM and early galaxy populations over the next decade are outstanding. Motivated by this, we review the current state of models of the IGM during reionization. We focus on a few key aspects of reionization-era phenomenology and describe: the redshift evolution of the volume-averaged ionization fraction, the properties of the sources and sinks of ionizing photons, along with models describing the spatial variations in the ionization fraction,
the ultraviolet radiation field, the temperature of the IGM, and the gas density distribution.}

\section{Introduction}
\label{intro}

Most of the volume of the universe, and much of the matter within it, lies in between the galaxies; this space is filled with diffuse gas known as
the intergalactic medium (IGM). The study of the IGM is important for a broad range of astrophysical topics including
studies of large-scale structure, galaxy formation, measurements of cosmological parameters, and in understanding the
overall history of our universe. First, the gas in the IGM mostly traces the underlying
matter density field -- which is apparently dominated by dark matter -- and the IGM hence probes the ``cosmic web''
of voids, sheets, filaments, and halos that characterize the overall distribution of matter in our universe on large
spatial scales \cite{Bond:1995yt}. Next, the gas out of which galaxies form {\em starts} in the IGM. As galaxies form, they 
impact the structure of the IGM by photoionizing it, heating it, and through galactic winds which enrich it with heavy elements (``metals''). Galactic winds also inject energy and momentum back into the IGM. These processes
then influence the formation of subsequent generations of galaxies, and so understanding the IGM is a pre-requisite for a complete
theory of galaxy formation. In addition, the properties of the IGM are ``nuisance'' parameters for several
probes of fundamental cosmological numbers: for example, parameter studies using the statistics of the Lyman-alpha (Ly-$\alpha$)
forest of absorption lines \cite{McDonald:2004eu} rely on a good understanding of the IGM, while the optical depth to electron scattering \cite{Zaldarriaga:2008ap} and the amplitude
of the patchy kinetic Sunayev-Zel'dovich effect \cite{Zahn:2005fn,McQuinn:2005ce} are nuisance parameters in studies of cosmic microwave background (CMB)
anisotropies and depend on IGM properties. Finally, and most relevant for
the present chapter, the early time properties of the IGM provide a {\em record of a key period in the history of
our universe, the Epoch of Reionization (EoR)} \cite{2013fgu..book.....L}.

The EoR marks the formation of early generations of stars, galaxies, and accreting black holes: as the first luminous
sources formed, they emitted ultraviolet radiation, ionizing ``bubbles'' of gas around
them. As ionized bubbles formed around neighboring sources, these regions grew and merged with each other, and
gradually filled  the entire volume of the universe with ionized gas.\footnote{The continuing output of UV
radiation from numerous generations of galaxies and accreting black holes has since kept the IGM in a highly ionized state to
the present day.}
This process is called
reionization and the time period over which it takes place is referred to as the EoR. The EoR is a key
time period in the history of our universe since it marks the formation of the first generation of complex
astrophysical objects, and because the reionization process impacted almost all of the baryons in the universe.
It is also a frontier topic in observational cosmology; observations across a wide range of wave-bands
are just starting to probe into the EoR and remarkable progress is expected in the next several years. In conjunction,
a great deal of theoretical work is being done to model the IGM during reionization; this work plays a crucial role
in interpreting current and future measurements, and in planning and forecasting
the sensitivity of upcoming observations.

The goal of this chapter is to describe some key features of IGM models during reionization, highlighting
some areas of recent progress as well as remaining challenges. The IGM modeling described in this chapter
is relevant for a broad range of current and upcoming data sets including: high redshift quasar absorption spectra \cite{Fan:2005es,Mortlock11}, narrow band surveys for
Lyman-alpha (Ly-$\alpha$) emitting galaxies (LAEs) \cite{Konno:2014fja}, measurements of the luminosity function of Lyman-break galaxies (LBGs) \cite{Bouwens:2014fua},
optical afterglow spectra of gamma-ray bursts \cite{Totani:2013nra}, improved measurements of CMB polarization \cite{Komatsu11}, 
small-scale CMB fluctuations \cite{Zahn12},
and redshifted 21 cm observations \cite{Parsons:2013dwa,Paciga:2013fj,Dillon:2013rfa,Yatawatta13}, among other probes.

In starting to think about models of the IGM during reionzation, it is interesting to first contrast this problem with the
case of modeling the {\em post-reionization IGM} at say $z \sim 2-4$. This is interesting because the post-reionization IGM
is subject to sharper empirical tests than presently possible during reionization (e.g. \cite{McDonald:2004eu,Palanque-Delabrouille:2014jca}). It is also a good starting point in
that cosmological hydrodynamic simulations provide a rather successful match to 
the main statistical properties of the $z \sim 2-4$ Ly-$\alpha$ forest \cite{Cen:1994da,Hernquist:1995uma,MiraldaEscude:1995bu,Hui:1996fh,Viel:2004bf}.\footnote{There are still, of course, interesting questions regarding
precisely how successful this basic model is and whether there are important missing ingredients. To name one prime example, HeII reionization should significantly impact the $z \sim 3$ IGM and
impact the thermal state of the IGM, and other properties (e.g. \cite{McQuinn:2008am,Compostella:2013zya}) in ways that are not included in most current Ly-$\alpha$ forest models.}
The main features of the successful post-reionization IGM model are summarized as follows. First, the IGM gas is highly photoionized, with only a small residual neutral fraction set by the balance between recombinations and photoionizations from
an approximately uniform UV radiation background. The UV radiation background is sourced by star-forming galaxies and quasars; current evidence suggests that star-forming galaxies produce
most of the UV radiation above $z \sim 3$ or so, and that quasars dominate at later times (e.g. \cite{1987ApJ...321L.107S,1999ApJ...514..648M,2008ApJ...688...85F}). 
Next, the temperature of the IGM
gas is determined mostly by the interplay between photoionization heating and adiabatic cooling from the expansion of the universe,
and should be close to a power law in the local gas density \cite{1994MNRAS.266..343M,Hui:1997dp}.\footnote{Although HeII reionization should impact this ``temperature-density'' relation, and lead to large-scale spatial fluctuations in
this relation \cite{McQuinn:2008am}.} Finally, on large scales, the IGM gas traces the overall distribution of matter
in the universe, while it is smoothed out on small scales by gas pressure gradients, i.e., by Jeans smoothing (e.g. \cite{Gnedin:1997td}). According to this model, the structure in Ly-$\alpha$
forest quasar absorption spectra trace fluctuations in the underlying line-of-sight density field with each spectrum providing a skewer through the cosmic web.
A key simplifying feature
of this successful model is that the UV radiation field is treated as spatially uniform. This is generally a good
approximation in the post-reionization universe because the mean free path to hydrogen ionizing photons is quite large, $\sim$ a few hundred co-moving Mpc at $z \sim 3$ (e.g. \cite{2009ApJ...705L.113P}). The approximation
of a uniform radiation field then allows modelers to mostly avoid detailed calculations of the radiative transfer of
ionizing photons through the IGM. 

However, the uniform UV radiation field approximation obviously breaks down close to and during the EoR, when radiative
transfer and a detailed modeling of the interplay between the ionizing sources and the IGM become essential. Indeed,
the overall timing of reionization should show strong spatial variations as some regions form galaxies and become
filled with ionized gas more rapidly than others. As a result, many of the properties of the IGM will fluctuate spatially
during reionization and an important challenge for IGM models is to account for the simultaneous variations in all of these quantities.

A wide variety of techniques are being employed to model the IGM during  reionization, including a range of different
schemes for approximately solving the radiative transfer equation (see \cite{2011ASL.....4..228T} for a recent review, and also the ``Cosmological Radiative Transfer Comparison Project" which tests and compares different methods, \cite{2009MNRAS.400.1283I}), often performed in a post-processing
step on top of evolved cosmological simulations. Another commonly used approach is the so-called ``semi-numeric'' modeling scheme \cite{Zahn:2006sg,2011MNRAS.411..955M}, based on the
excursion set formalism \cite{Furlanetto:2004nh,1991ApJ...379..440B}. 
In this chapter, we focus mostly on a few important aspects of {\em reionization-era IGM phenomenology} rather than on the specifics of different 
radiative transfer calculations and semi-numeric models; we refer the reader to the above articles for more information regarding the details of these techniques. 

The outline of this chapter is as follows. In \S \ref{sec:xi_av}, we discuss an approximate model for the volume-averaged
ionization fraction of the IGM and its redshift evolution. This in turn depends on the properties of the sources and sinks of ionizing
photons, which are further discussed in \S \ref{sec:sources_further} and \S \ref{sec:sinks}, respectively. We then turn to consider
spatial variations in the ionized fraction and the size distribution of the ionized regions during reionization in
\S \ref{sec:bubbles}. After this, we discuss models that describe the spatial variations in the UV radiation background (\S \ref{sec:uvb_fluc}) and the temperature of the IGM after reionization (\S \ref{sec:temp_igm}). \S \ref{sec:gas_pdf} describes models of
the gas density distribution. We conclude in \S \ref{sec:conclusion}, briefly summarizing the present state of IGM models along with some future prospects.

\section{The Volume-Averaged Ionization Fraction and its Redshift Evolution}
\label{sec:xi_av}

We start by considering a simple model for the average ionization history of the universe, with the aim of providing
an approximate description of how reionization proceeded over cosmic time. Empirical constraints on the ionization history
can determine: when in our cosmic history did the reionization process begin and how long did it take for the universe to become
filled with ionized gas? The answers to these questions are, of course, intimately tied to the properties of the ionizing sources and to the clumpiness 
of the intergalactic gas. 

The mean free path of ionizing photons propagating through the neutral IGM during reionization is quite short. As a result,
the ionization state of the IGM should be well-described as a two-phase medium, and consist 
of highly-ionized regions intermixed with gas in a mostly-neutral phase.\footnote{In principle, a two-phase medium might be a poor description if the ionizing sources have a very hard spectrum, but this possibility seems unlikely given existing observations \cite{McQuinn:2012bq}.} A key quantity of interest is the
fraction of the IGM volume in each of the ionized and neutral phases as a function of time.  A major goal is to
robustly extract the redshift evolution of this quantity from upcoming observations.
Here we focus on a simple but approximate model for describing the redshift evolution of the volume-averaged
ionization fraction. This illustrates what we hope to learn from future observational constraints on the ionization
fraction, while revealing the key ingredients involved in the
associated modeling.

An approximate equation describing the redshift evolution of the ionized fraction is \cite{1987ApJ...321L.107S,Haiman:1996ht,1999ApJ...514..648M}:
\beq
\frac{d \avg{x_i}}{dt} = \frac{d\left(n_\gamma/n_H\right)}{dt} - \frac{\avg{x_i}}{\bar{t}_{\rm rec}}.
\label{eq:xi_ev}
\eeq
Here $\avg{x_i}$ denotes the volume-averaged ionization fraction, $n_\gamma/n_H$ is the average number of ionizing
photons per hydrogen atom and $\bar{t}_{\rm rec}$ is the average recombination time of gas in the IGM. 
This equation reflects the competition between
photoionizations (first term on the right-hand side) and recombinations (second term on the right-hand side). 

Since the average time between recombinations in the reionization-era IGM is fairly long, 
the ionization fraction during the EoR mostly reflects
the cumulative ionizing photon output of {\em all of the luminous sources} that have turned-on up to that time. This is in contrast
to measurements of the galaxy luminosity function at high redshift, which are sensitive to only those sources above some luminosity
limit (and at a given instant in cosmic time). In fact, present evidence suggests that {\em most} of the ionizing photons at $z \gtrsim 6$ are produced by sources below
present detection limits (see \S \ref{sec:sources_further} below).  This feature highlights one of the main strengths of using the properties of the IGM to learn about
the ionizing sources: the ionization state of the IGM reflects the combined influence of all of the ionizing sources, and thereby complements measurements
of galaxy counts which are able (at current sensitivities) to detect only relatively bright objects.

It is worth keeping in mind some of the assumptions and limitations inherent in Eq.~\ref{eq:xi_ev}. First, the average recombination time depends on the
clumpiness of the ionized gas in the intergalactic medium. This depends, in turn, on the details of reionization itself. For example, if the ionizing
sources have an especially hard spectrum the ionization fronts will penetrate more deeply into dense clumps, where the density and recombination
rate are higher. Second, some care is required in deciding which regions to include in the averaging used to define the clumping factor. For example, self-shielded
highly neutral regions should not be included in the averaging. In addition, note that
the escape fraction $f_{\rm esc}$ -- that is, the fraction of ionizing photons that escape the host halo and make it into the IGM (see \S \ref{sec:sources}) -- implicitly incorporates the impact of recombinations within the host halos of the ionizing sources, and so including
regions within the halos of ionizing sources would (mostly) be double-counting.\footnote{Note that recombinations that occur following photoionizations from an exterior source would not involve
double-counting, hence the parenthetical ``mostly'' remark here.} 
Third, Eq.~\ref{eq:xi_ev} ignores spatial correlations between the sources and sinks of ionizing photons. Fourth, this equation effectively assumes that ionizing photons are absorbed instantaneously, and ignores
redshifting effects. Finally, although an ensemble averaged
clumping factor suffices to determine the average ionization fraction, in some contexts it may be important that the clumping factor likely has large
spatial fluctuations.

\subsection{The Source Term}
\label{sec:sources}

The first ingredient in the above equation is the source term, describing the rate of production of ionizing photons. In order to develop some intuition here,
let us first consider the simplest possible model for the {\em cumulative} number of ionizing photons emitted (per hydrogen atom) by the ionizing sources.
In particular, let us assume that each dark matter halo above some minimum mass $M_{\rm min}$ hosts a galaxy and that the cumulative output of ionizing photons
per hydrogen atom is:
\beq
\frac{n_\gamma}{n_H} = f_{\rm esc} f_\star N_{\gamma} f_{\rm coll}(>M_{\rm min}) = \zeta f_{\rm coll}(M>M_{\rm min}).
\label{eq:photon_count}
\eeq
This equation expresses the ionizing photon budget as a product of several uncertain factors, which one might loosely refer to as ``reionization's Drake Equation''.\footnote{The
Drake Equation describes the likelihood of extraterrestrial life as the product of several uncertain factors.}
Here $f_{\rm esc}$ is the fraction of ionizing photons that escape the host halo and make it into the IGM. The escape of ionizing radiation from each
individual galaxy likely varies significantly with propagation direction, time, and scale. In addition, it depends on the detailed spatial distribution of the gas, stars, and dust in the
interstellar medium of the galaxy as well as the distribution of ``circumgalactic'' gas in the host halo.
The above description sweeps these complexities into a single parameter, which should be thought of as a global
average over time, direction, and host galaxy properties. Note that the above formula ignores any explicit dependence of the escape fraction and other quantities on host halo mass, although it is straightforward to allow additional mass dependence. Next, $f_\star$ describes the fraction of the baryons in the halo
that have been converted into stars; the above equation implicitly assumes that each halo contains the universal cosmic baryon fraction. The quantity $N_{\gamma}$ is the number of ionizing photons produced per baryon converted into stars. The ionizing photons are produced by O and B stars (with high surface temperatures and short lifetimes), and perhaps by metal-free Pop III
stars as well. The ionizing photon yield, $N_{\gamma}$, is hence sensitive to the Initial Mass Function (IMF) and metallicity of these stellar populations; the IMF determines the fraction of stellar mass that is
incorporated into the massive, ionizing photon producing, stars. 
Finally, $f_{\rm coll} (> M_{\rm min})$ is the fraction of matter that has collapsed into halos above some minimum mass $M_{\rm min}$. Note that for simplicity this equation neglects an order unity correction coefficient (to account for the presence of helium, e.g. \cite{2013fgu..book.....L}).

Here $M_{\rm min}$ is meant to represent a plausible minimum host halo mass above which gas can cool and condense to form stars.  The quantity $\zeta$ in Eq. \ref{eq:photon_count} 
then describes the efficiency at which matter collapsing into galaxy hosting halos produces ionizing photons.\footnote{Note that in some work $\zeta$ is defined differently than here, and incorporates a
factor describing the average number of recombinations per hydrogen atom. Here recombinations are treated in the sink term of Eq. \ref{eq:xi_ev}.} 
As gas falls into a collapsing dark matter halo, it shocks and heats to the virial temperature with $k_B T_{\rm vir} = \mu m_p G M/2 r_{\rm vir}$, where $M$ is the total gravitating mass
of the collapsed halo of radius $r_{\rm vir}$, $\mu$ is the mean mass per particle in units of the proton mass, and the equality reflects the virial balance between the kinetic and gravitational energies in the collapsed halo.\footnote{Note, however, that much of the gas in lower mass halos may fall into a growing galaxy along filaments in ``cold mode accretion'' flows without shocking to the halo virial temperature \cite{Birnboim:2003xa,Keres:2004cq}.} An important mass scale is then the mass at which the virial temperature reaches $T_{\rm vir} = 10^4$ K, because primordial atomic gas cooler than this is unable to cool, fall to the center of the
halo, and ultimately fragment and form stars.\footnote{At higher temperatures, collisions excite atoms to energy levels above the ground state: the excited atoms quickly decay and emit photons,
some of which escape the halo and cool the gas. Below $10^4$ K, the gas is not hot enough to excite hydrogen atoms above the ground state.} Cooling by molecular and metal lines may allow the gas
to cool in smaller halos, although molecular gas is fragile and easilly dissociated. The halo virial temperature and mass, at collapse redshift $z$, are related by: (e.g. \cite{Barkana:2000fd}):
\beq
T_{\rm vir} = 1.1 \times 10^4 K \left[\frac{M}{10^8 M_\odot}\right]^{2/3} \left[\frac{1+z}{7}\right].
\label{eq:tvir}
\eeq
A first rough estimate of the minimum galaxy-hosting halo mass is then $M_{\rm min} \sim 10^8 M_\odot$, as this corresponds roughly to the halo mass above which the gas can cool by emitting in atomic lines at the redshifts of interest. 

The minimum mass should, however, be influenced by numerous forms
of feedback from the galaxy formation process: supernova winds can expel gas from small host halos (e.g. \cite{Dekel86}), photoionization can heat gas sufficiently to prevent it from failling into low mass halos (e.g. \cite{Efstathiou92,Thoul96,Barkana:1999apa,Gnedin:2000uj,Dijkstra04,Shapiro:2003gxa,Okamoto:2008sn,Sobacchi:2013wv}), and
perhaps remove gas from existing galaxy-hosting halos, while UV radiation can also impact the cooling rate of collapsing gas by dissociating molecules \cite{Haiman:1995jy} and photoionizing atomic gas.
This quantity should be time dependent and vary spatially since photoionization heating will impact only regions where galaxies have turned on and produced a significant amount
of UV radiation. It may not, however, impact all galaxies forming within ionized regions: some of the gas in newly forming galaxies may have fallen into the host halo long before and reached a high enough density to self-shield from photo-ionizing radiation before being exposed to this radiation \cite{Dijkstra04}. It is also possible that supernova feedback is sufficiently strong to mostly overwhelm the impact of photoionization heating. In this case, since supernova feedback depends on local galaxy-scale physics, spatial variations in the timing of reionization may {\em not} significantly modulate the efficiency of the galaxy formation process. 
Also note that the impact of feedback effects should depend most directly on the depth of the halo potential well, and so it may be better to consider a minimum virial temperature for galaxy hosting halos, rather than a minimum host halo mass. Here we will, however, stick to using $M_{\rm min}$ as our parameter.

\subsection{Clumping Factors}
\label{sec:clumping}

The rate at which ionized gas in the IGM recombines scales as density {\em squared}, and so the volume-averaged recombination rate depends on the so-called clumping factor
$C = \avg{\rho_g^2}_{\rm ionized\ IGM}/\avg{\rho_g}^2$. Assuming the case-B recombination rate here\footnote{The case-B recombination rate excludes recombinations directly
to the ground state in calculating the total recombination rate of the gas. The rationale here is that (direct) recombinations to the ground state produce an additional ionizing photon; if these
photons are quickly absorbed nearby, ``on-the-spot'', then these recombinations have no net effect on the ionization state of the gas.} and a temperature of
$T=2 \times 10^4 K$, the average time between recombinations is:
\beqa
\bar{t}_{\rm rec} = 0.93 {\rm Gyr} \left[\frac{3}{C}\right] \left[\frac{1+z}{7}\right]^{-3} \left[\frac{T_0}{2 \times 10^4 K}\right]^{0.7}.
\label{eq:trec}
\eeqa

As we emphasized previously, the clumping factor must -- at some level -- depend on the nature of reionization itself: the clumpiness of {\em the ionized gas} is the relevant quantity here, and precisely which regions
are ionized depends on the spectrum of the ionizing sources, the intensity of the UV radiation field incident on dense regions, and other details of the reionization process. In addition, the interplay between
radiative transfer and hydrodynamics should also be important in this problem: as ionizing photons penetrate into a dense region and heat the interior gas to temperatures greater than the virial
temperature of the host halo, a photo-evaporative flow is produced and gas gradually escapes the halo \cite{Shapiro:2003gxa}.  

Despite these challenges, a great deal of progress has been made recently by using small-scale hydrodynamic simulations to measure clumping factors; these results can then be incorporated into
large-volume reionization simulations through sub-grid modeling and into analytic calculations (using Eq. \ref{eq:xi_ev}, for example). In one study, \cite{Pawlik:2008mr} used SPH simulations with a uniform
UV background radiation field and considered the impact of photo-heating in the optically thin approximation. These authors emphasize that photo-heating exerts a positive feedback on reionization -- by
reducing the clumpiness of the IGM -- as well as the more widely appreciated negative feedback (from raising $M_{\rm min}$). In order to separate recombinations in the IGM from those
in the ISM of a galaxy (which are mostly accounted for in the escape fraction), and to approximately account for self-shielded gas, these authors consider the clumping factor of gas beneath
various overdensity thresholds. For a threshold gas overdensity of $\Delta = \rho_g/\avg{\rho_g} = 100$, this study finds $C=3$ at $z=6$ for gas reionized at $z_r \gtrsim 9$. Shortly after
a gas element is ionized, the clumping factor is higher than this (see their Fig. 5), because it takes some time for the gas to respond to prior photo-heating.\footnote{Note that this timescale
may be underestimated owing to the optically thin approximation adopted in this work. In the optically thin limit, the response time is roughly the sound crossing time, $L_J/c_s$, where $L_J$
is the Jeans length (Eq. \ref{eq:ljeans}) and $c_s$ is the sound speed in the reheated gas.}  After this ``response time'' passes, the clumping factor depends weekly on redshift. 
Note, however, that it is challenging to fully resolve the small-scale structure in the gas distribution before it has had time to relax, and so the finite resolution of the simulations in the above study
may lead to an underestimate of the early time clumping factors \cite{Emberson13}. 
\cite{McQuinn:2011aa} also calculate the clumping factor
from SPH simulations, treating the radiative transfer in a post-processing step, and thereby explicitly accounting for the self-shielding of dense regions (although their post-processing approach does not
capture the coupling between the hydrodynamics and radiative transport). These authors find $C=2-3$ at $z=6$ for gas that reheated at $z_r =10$. Another earlier work (\cite{2007ApJ...657...15K}),  uses full small-scale radiation hydrodynamic
simulations of reionization; this work uses several different weighting schemes to compute the averages that enter into the clumping factor calculations. This work also
shows a sizeable scatter in the locally-estimated clumping factors. It is worth mentioning that all of these studies ignore the impact of ``pre-heating'' from early X-rays \cite{Oh:2003pm,Furlanetto:2006tf}, which may heat the gas up to temperatures as large as $T \sim 1,000$ K significantly before reionization, reducing the clumping factor at early times and the time scale for the gas to relax after subsequent heating.

\subsection{Model Reionization Histories}
\label{sec:xi_model}

\begin{figure}[b]
\sidecaption
\includegraphics[scale=0.65]{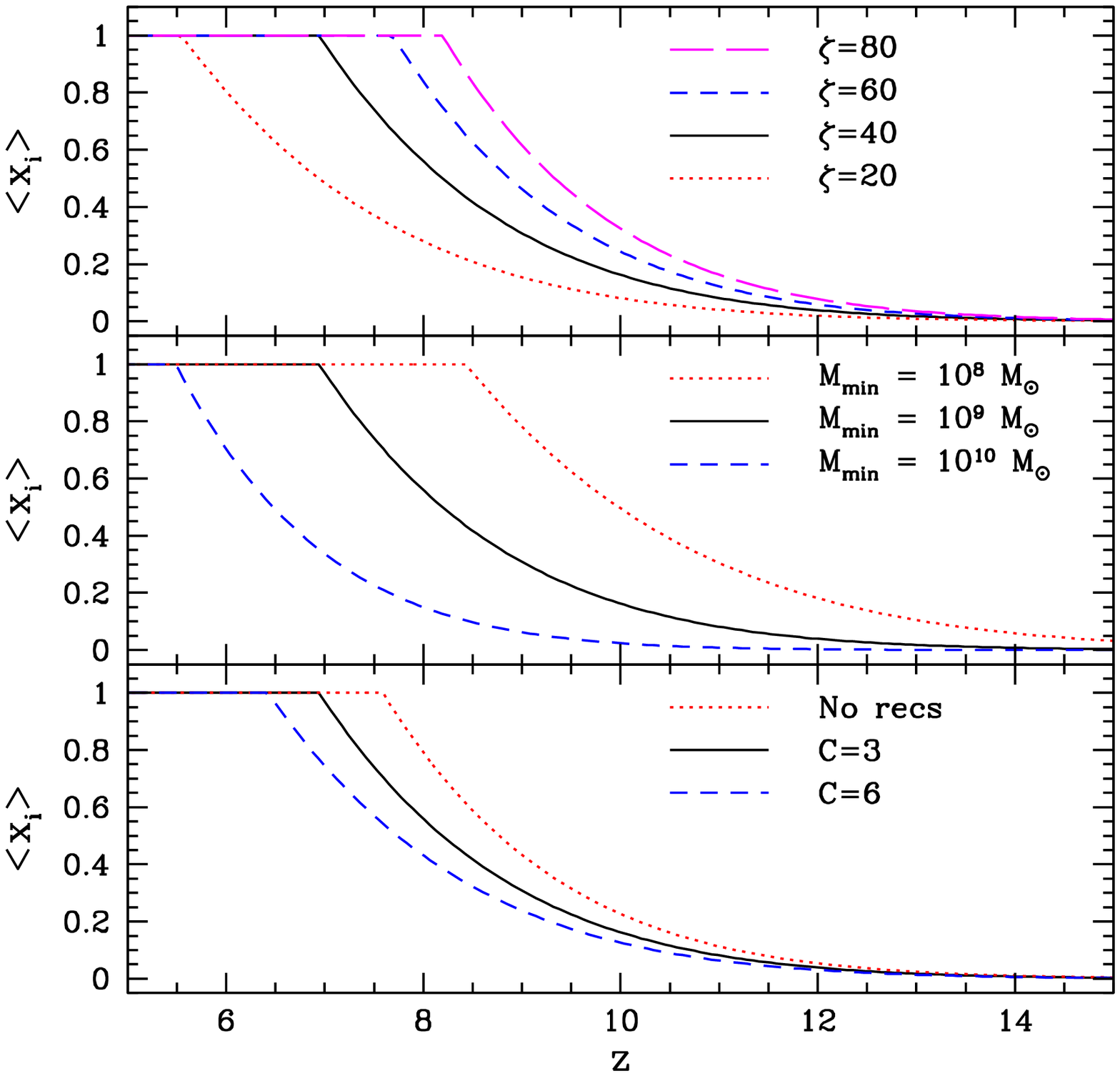}
\caption{Illustrative models of $\avg{x_i}(z)$. {\em Top panel}: The impact of variations in the ionizing efficiency parameter on the ionization history of the universe. {\em Middle panel}: Results for varying values of the minimum galaxy-hosting halo mass.  {\em Bottom panel}: The curves here vary the clumping factor of the IGM, or equivalently the average recombination time in the
ionized IGM.}
\label{fig:xi_vary}
\end{figure}

In summary, let us consider solutions to Eq. \ref{eq:xi_ev} for plausible values of the ionizing efficiency, $\zeta$, the minimum mass, $M_{\rm min}$, and the clumping factor, $C$.
Inserting a typical set of fiducial, but uncertain, numbers gives:
\beqa
\zeta = 40 \left[\frac{f_\star}{0.1}\right]  \left[\frac{f_{\rm esc}}{0.1}\right] \left[\frac{N_\gamma}{4000}\right]
\label{eq:zeta_fid}
\eeqa
 for the ionizing efficiency parameter, $\zeta$, while $M_{\rm min} = 10^9 M_\odot$ and $C=3$ are plausible numbers for the minimum host halo mass and clumping factor, respectively.
The fiducial value of $M_{\rm min}=10^9 M_\odot$ here is a bit higher than the atomic cooling mass described in Eq. \ref{eq:tvir}; this is intended to roughly account for the negative feedback from photoionization heating
and supernova feedback. Adopting a constant value for the minimum host halo mass rather than one that varies in time and spatially is unlikely to be realistic in detail, but we are just interested in rough estimates here.
 
 Fig. \ref{fig:xi_vary} shows the redshift evolution of the volume-averaged ionization fraction for this case, and also illustrates the impact of parameter variations around this model. 
 The top panel of the figure shows
 the ionization history for varying $\zeta$ between $\zeta=20$ and $\zeta=80$ (in steps of $20$), while fixing $M_{\rm min} = 10^9 M_\odot$ and $C=3$. Note that this range is not meant to span the {\em full} range of possible values, but only
 to give some sense for the dependence of the ionization history on this parameter. This range in $\zeta$ might correspond, for example, to  lowering and raising the escape fraction from
 its fiducial value by a factor of two.  As one simple description of the resulting ionization history, these values of $\zeta$ lead to the following range of redshifts for reionization to ``complete'' (i.e., the $z=z_{\rm end}$ at which $\avg{x_i}$ reaches
 unity): $\zeta=(20, 40, 60, 80)$ gives $z_{\rm end}=(5.6,6.9,7.7,8.2)$. For each ionization history, we can additionally calculate the probability that a CMB photon Thomson scatters off a free electron during
 and after reionization. Current constraints on the optical depth to electron scattering, $\tau_e$, come from WMAP measurements of the 
 E-mode polarization power spectrum combined with Planck temperature
 anisotropy data and give $\tau_e = 0.089^{+0.12}_{-0.14}$ \cite{Ade:2013zuv,Bennett:2012zja}. For reference, the models with $\zeta=(20,40,60,80)$ give $\tau_e=(0.051,0.063,0.070,0.075)$ assuming helium is singly ionized along with hydrogen and ignoring the expected small
 increase in this value following HeII reionization. Current measurements hence prefer the higher ionizing efficiency models. In the middle panel, we show the impact of varying $M_{\rm min}$ across
 $M_{\rm min} = 10^8 M_\odot, 10^9 M_\odot$, and $10^{10} M_\odot$, while fixing $\zeta=40$ and $C=3$. Reionization starts earlier for the smaller value of $M_{\rm min}$, while it is delayed for
 the larger $M_{\rm min}$. This reflects the hierarchical nature of structure formation in Cold Dark Matter (CDM) cosmologies: small halos collapse first, and larger halos are built up subsequently from
 the merging of smaller systems. Quantitatively, we find $(z_{\rm end}, \tau_e) = (8.4,0.084); (6.9,0.063);(5.5,0.045)$ for $M_{\rm min}=10^8 M_\odot$, $10^9 M_\odot$, and $10^{10} M_\odot$, respectively.
 Finally, the bottom panel shows the impact of varying the clumping factor and hence the average time between recombinations in the ionized IGM (while fixing $\zeta=40$ and $M_{\rm min}=10^9 M_\odot$):
 this gives $(z_{\rm end},\tau_e)=(6.4,0.059);(6.9,0.063);(7.6,0.069)$ for $C=6, 3$, and $C=0$,  respectively. (The $C=0$ case gives the ionization history in the absence of recombinations).

Fig. \ref{fig:xi_vary} demonstrates that a fairly broad range of ionization histories are possible, even within the context of this simple model. In reality, the efficiency parameter likely has some mass and
redshift dependence, while the minimum mass should depend on redshift and vary spatially as reionization proceeds, as should the clumping factor. Nonetheless, if upcoming measurements
can place constraints on the reionization history, $\avg{x_i(z)}$, this  will provide a powerful record of the cumulative impact of all previous generations of ionizing sources.  This information can
then be combined with direct measurements of the UV luminosity density of star-forming galaxies over cosmic time to further reveal the nature of these early galaxy populations.

\section{Guidance from Existing Observations: The Sources of Ionizing Photons}
\label{sec:sources_further}

The model of the previous section is a useful start, but here we will delve a little more deeply and give a brief overview of the current status of {\em empirical} constraints on the properties
of the ionizing sources.
A great deal of observational progress has been made recently, enabled in large part by the Wide-Field Camera 3 (WFC-3) onboard the Hubble Space Telescope (HST), in measuring high redshift UV galaxy luminosity
functions (e.g \cite{2010ApJ...709L.133B,2010MNRAS.409..855B,Robertson:2013bq,2014arXiv1410.5439F}). These measurements provide important empirical guidance regarding the properties of the ionizing sources, which can be used to inform reionization models. For the most
part, galaxy properties are hard to capture in first-principles simulations -- especially those aiming to capture the large volumes relevant for reionization studies -- and so modelers rely
on sub-grid prescriptions to describe the ionizing sources. Hence the UV luminosity function measurements provide both general insight into the properties of the ionizing sources, and can help
guide these sub-grid models.

\begin{figure}[b]
\sidecaption
\includegraphics[scale=0.65]{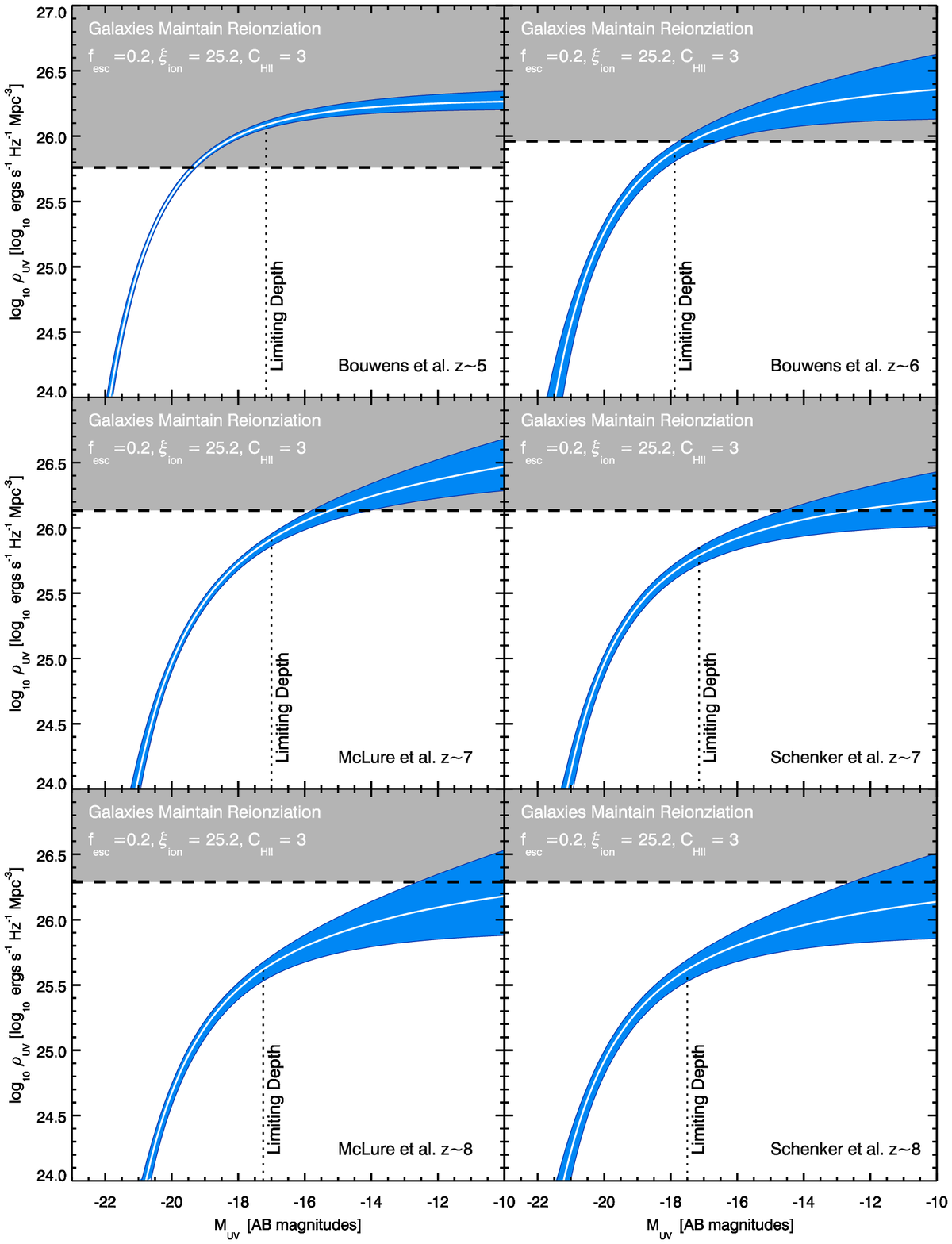}
\caption{Empirical constraints on the UV luminosity density produced by star-forming galaxies. In each panel the blue-shaded band shows the estimated UV luminosity density (at a restframe wavelength of $1500\text{\AA}$), as a function of limiting magnitude, as inferred from Schechter-function fits to the measured galaxy luminosity functions at various redshifts. The dotted line shows the limiting magnitude of the luminosity function measurements; the results to the right of this limit are extrapolations based on the faint end slope of the Schechter function fits. The upper shaded region (above the horizontal
dashed line) in each panel shows the ``critical'' UV luminosity density. This critical value is set so that the corresponding ionizing emissivity is just sufficient to balance recombinations and maintain the ionization state of the IGM. This band
assumes $f_{\rm esc} = 0.2$, a clumping factor of $C=3$, and the value of $\xi_{\rm ion}$ discussed in the text. The different panels show measurements at various redshifts, as labeled. From \cite{Robertson:2013bq}. }
\label{fig:uv_robertson}
\end{figure}

The luminosity function measurements directly determine the UV luminosity density above some limiting magnitude, which are well-fit by Schechter functions \cite{1976ApJ...203..297S}.
In order to infer the total rate per unit volume at which star-forming galaxies
produce ionizing photons, additional assumptions are required. Specifically, this conversion depends on the escape fraction of ionizing photons, the UV spectral shape of the ionizing sources,
and  assumptions about the luminosity density in sources below the limiting magnitude of the observations. One may write \cite{Robertson:2013bq}:
\beqa
\dot{n_\gamma} = f_{\rm esc} \xi_{\rm ion} \rho_{\rm UV},
\label{eq:nphot}
\eeqa
where $\dot{n_\gamma}$ denotes the number of ionizing photons per co-moving volume produced per unit time, $f_{\rm esc}$ is the escape fraction, $\xi_{\rm ion}$ quantifies the number of ionizing
photons produced per erg per second per Hz of UV luminosity emitted at a restframe wavelength of $1500\text{\AA}$, and $\rho_{\rm UV}$ denotes the UV luminosity density (at $1500\text{\AA}$) in units 
of ${\rm ergs}~ {\rm s}^{-1} {\rm Hz}^{-1} {\rm Mpc}^{-3}$. \cite{Robertson:2013bq} explore a variety of Bruzual \& Charlot (\cite{Bruzual:2003tq}) stellar population synthesis models, and find that a range of values for $\xi_{\rm ion}$ are
consistent with the observed UV spectral slopes, adopting the plausible value of $\xi_{\rm ion} = 10^{25.2} {\rm ergs}~{\rm s}^{-1} {\rm Hz}^{-1}$. In this case, it is useful to note that:
\begin{align}
\rho_{\rm UV} = &1.6 \times 10^{26}~ {\rm ergs}~{\rm s}^{-1} {\rm Hz}^{-1} {\rm Mpc}^{-3} \left[\frac{0.2}{f_{\rm esc}}\right] \left[\frac{\dot{n_\gamma}}{3~ {\rm photons/hydrogen~ atom/Gyr}}\right] \nonumber \\
& \times \left[\frac{10^{25.2} {\rm ergs}~{\rm s}^{-1} {\rm Hz}^{-1}}{\xi_{\rm ion}}\right].
\label{eq:rho_uv}
\end{align}
The ionizing emissivity here has been normalized to $\dot{n_\gamma} \sim 3$ photons per atom per Gyr. This is a characteristic value because it is comparable to the value $\dot{n_\gamma} \sim C \alpha n_e = 2$ photons per atom per Gyr required to {\em maintain} the ionization of the IGM at $z=6$, assuming the case-A recombination coefficient\footnote{See \cite{MiraldaEscude:2002yd} for a discussion regarding why the case-A recombination coefficient may be more appropriate than the case-B rate in this context.}   at a temperature of $T=2 \times 10^4$ K, and $C=3$. In the above
equation we have set $f_{\rm esc}=0.2$, as in \cite{Robertson:2013bq}, because escape fractions of this order appear necessary for galaxies to reionize the universe and for these sources to simply maintain
the ionization at later times. This
escape fraction seems required unless either early stellar populations produce a surprisingly large number of ionizing photons per baryon converted into stars, or else there is an abundant
faint source population above that predicted by even generous extrapolations down the faint end of the Schechter function fits. 

The observed galaxy UV emissivity at redshifts $z \sim 5-8$, compiled from various recent UV luminosity function measurements in the literature \cite{2007ApJ...670..928B,2013MNRAS.432.2696M,2013ApJ...768..196S} by \cite{Robertson:2013bq}, is shown in Fig. \ref{fig:uv_robertson}. The blue
shaded band in each panel shows the range in luminosity density allowed by Schechter function fits to the observations. In each panel, the vertical dotted line at $M_{\rm UV} = -17$
indicates the limiting depth of the measurements, and so the portions of the blue band that lie to the right of the vertical dotted lines are extrapolations to fainter luminosities than currently observed.
Evidently, the luminosity functions at these redshifts rise quite steeply toward the faint end, suggesting that numerous individually dim sources may play a dominant
role in reionizing the universe. For comparison, the grey region shows the value of the UV luminosity density required simply to maintain the ionization of the IGM for $C=3$, $f_{\rm esc}=0.2$ and the above $\xi_{\rm ion}$,
as described in Eq. \ref{eq:rho_uv}.\footnote{The values in the figure are slightly different than in Eq. \ref{eq:rho_uv} because \cite{Robertson:2013bq} adopt the case-B recombination coefficient, while
the equation here assumes the case-A recombination coefficient.}  It is interesting to note that even for $f_{\rm esc} = 0.2$, galaxies above current survey limits are only barely able to maintain the ionization of the IGM
at $z =6$, while at higher redshifts a significant contribution from lower luminosity sources is required to maintain the ionization of the IGM. If the escape fraction is smaller, or if the clumping factor is larger than assumed here, an even more sizable contribution is required from the low luminosity galaxies. 

Another independent approach for estimating the ionizing emissivity involves measurements of the average level of absorption in the Ly-$\alpha$ forest (e.g., \cite{Bolton:2007b,2008ApJ...688...85F}). The measured mean absorption (or equivalently
mean-transmitted flux, $\avg{F}$) can be used, in conjunction with numerical simulations of the IGM, to infer the intensity of the UV radiation background. The inferred amplitude of the UV background
can then be combined with an estimate of the mean free path to ionizing photons (which can be extracted from the column density distribution of absorbers in the forest \cite{2010ApJ...721.1448S} or by stacking absorption
spectra near the wavelength of the Lyman limit, i.e. near a rest-frame wavelength of $\lambda = 912\text{\AA}$ -- see \cite{Worseck:2014fya}), to determine the ionizing emissivity. One benefit of this approach is that it is sensitive to the total intensity of the UV background radiation,
and hence to the combined influence of all of the ionizing sources. This is in contrast to the UV luminosity function measurements, which are only currently sensitive to bright sources. In addition, one can
infer the ionizing emissivity with this technique without making assumptions about the escape fraction of ionizing photons. However, the results do depend somewhat on the model of the IGM; for example,
the inferences depend on the temperature and density distribution of absorbing gas in the IGM, and this approach also requires an accurate measurement of the mean free path to ionizing photons. Interestingly, near $z \sim 5-6$, the
ionizing emissivity inferred from this approach is comparable to the emissivity mentioned above, $\sim$ a few photons per atom per Gyr (e.g. \cite{MiraldaEscude:2002yd,Bolton:2007b,Kuhlen:2012vy}), although \cite{Becker:2013ffa} recently argued for a bit larger an emissivity.

The relatively low value of the emissivity of ionizing photons inferred from this data has been used to argue that reionization is ``photon starved'' \cite{Bolton:2007b}. Note that, accounting for recombinations, it should take
a few photons per atom to complete reionization and that the age of the universe near $z \sim 6$ is close to one Gyr. An emissivity of a few photons per atom per Gyr then implies that sources
emitting at this rate, over the entire age of the universe, are just capable of completing reionization by $z \sim 6$. The simplest explanation for the inferred $z \sim 5-6$ emissivity is then that
reionization was a gradual process, and that it completed near $z \sim 6$, rather than at significantly higher redshift. An alternative possibility is that the ionizing emissivity was higher above $z \gtrsim 6$; in
this case, reionization may have completed at higher redshift, and then the emissivity may have fallen to match the values inferred from the $z \sim 5-6$ Ly-$\alpha$ forest and the galaxy luminosity function measurements.  See e.g. \cite{Alvarez12} for  a model of this type; in their model, the declining emissivity is driven by photoionization feedback and a decreasing escape fraction. 

In any case, these measurements of the UV luminosity functions and of the ionizing emissivity after reionization can be used to limit the range of possible models for the ionizing sources. Specifically, the low ionizing
emissivity suggested by these observations can be used to argue against  some of the prescriptions for the ionizing sources used in the simulation literature; the ionizing emissivity
in these prescriptions often grows rapidly toward decreasing redshift and these models may therefore overshoot the post-reionization emissivity constraints (as pointed out in e.g. \cite{Choudhury:2008aw,Mesinger12,Sobacchi:2014rua}). Further work in this direction should help improve reionization models.

\section{The Sinks of Ionizing Photons}
\label{sec:sinks}

In addition to describing the {\em sources} of ionizing photons, reionization models must capture the {\em sinks} of ionizing photons.
While the clumping factor approach is adequate for rough estimates of the volume-averaged ionization fraction (Eq. \ref{eq:xi_ev}), it does not address in which environments
the absorptions take place, and is therefore entirely insufficient for understanding the spatial distribution of the ionized gas. We hence turn now to consider the sinks of ionizing photons more
closely.  

Especially towards the end of reionization, many ionizing photons will be absorbed in dense ``clumps'' and these systems will play
a key role in moderating the growth of ionized regions and in setting the mean free path to ionizing photons \cite{MiraldaEscude:1998qs,Furlanetto:2005xx}. These dense clumps
are observed at slightly lower redshifts as Lyman-limit systems in quasar absorption spectra. One of the challenges involved in
modeling reionization now becomes apparent: a large {\em dynamic range} is required to resolve the dense clumps, which
play an important role as sinks of ionizing photons, while simultaneously capturing a representative sample of the ionized regions, which are likely large. Resolving the sources of ionizing photons
in a large volume is also, of course, difficult.
For instance, if $\Delta x \sim 10$ kpc-scale spatial resolution is required in a $L_b \sim 100$ Mpc box (which seems like fairly
minimal requirements given that the ionized regions may be tens of co-moving Mpc in size and given plausible estimates of dense clump size, Eq. \ref{eq:ljeans}), this necessitates simulating $(L_b/\Delta x)^3 \sim 10^{12}$ resolution elements. In addition, the interplay between radiation and gas dynamics likely plays an important role in understanding the
precise impact of the sinks of ionizing photons, as we discussed earlier in the context of clumping factors (e.g. \cite{Shapiro:2003gxa}). Given these challenges, it seems that the best approach at present is to incorporate these sinks into
large-volume reionization simulations using a sub-grid model informed by smaller-volume simulations.

It is useful to start with a rough estimate of which regions in the IGM are able to self-shield in the presence of photoionizing
UV radiation. Here it is helpful to relate the neutral hydrogen column density -- which determines whether a clump can self-shield --
and the physical overdensity of a clump. Here we follow the Appendix of \cite{Furlanetto:2005xx}, which is in turn based partly on
the work of \cite{Schaye:2001me}, who aimed to elucidate a few key features of the $z \sim 3$ Ly-$\alpha$ forest.
The starting point here is to assume that the typical size of absorbing regions is on the order of the local Jeans length. The local Jeans
length is the length scale over which the sound-crossing time $t_{\rm sound} = L/c_s$ is equal to the dynamical time
$t_{\rm dyn} = 1/\sqrt{G \rho}$ where $c_s$ is the sound speed in the gas, and $\rho$ is the total matter density. In evaluating the
sound-speed we approximate the gas as isothermal, and adopt a mean mass per particle in the gas of $\mu m_p = 0.61 m_p$ (appropriate
for primordial gas with highly ionized hydrogen and singly-ionized helium).
Equating the sound crossing time with the dynamical time 
and solving for the $L=L_J$ at which the equality holds (assuming that the fraction of absorber mass in gas is equal to the cosmic mean baryonic fraction):
\beqa
L_J = 50\ {\rm proper}\ {\rm kpc} \left[\frac{\Delta}{1}\right]^{-1/2} \left[\frac{T}{10^4 K}\right]^{1/2} \left[\frac{1+z}{7}\right]^{-3/2}.
\label{eq:ljeans}
\eeqa
Here $\Delta=\rho_g/\avg{\rho_g}$ is the gas density in units of the cosmic mean, and so $\Delta=1$ refers to gas at the cosmic mean density.

The neutral hydrogen column density across this absorber is then $N_{\rm HI} \approx n_{\rm HI} L_J$. For the moment, we further assume photoionization equilibrium with an incident
UV radiation field of photoionizing intensity $\Gamma_{\rm HI}$ (with units of $s^{-1}$); photoionization equilibrium dictates
that $\Gamma_{\rm HI} n_{\rm HI} = \alpha n_e n_p$. In what follows, we adopt the case-A recombination coefficient with
$\alpha_A \approx 4.2 \times 10^{-13}$ cm$^3$ s$^{-1}$ at $T=10^4$ K \cite{Hui:1997dp}. In this case, there is a one-to-one relation between column density
and overdensity, and an estimate of the overdensity at which the gas becomes self-shielding follows from considering when
$N_{\rm HI} \approx 1/\sigma_{\rm HI}$, with $\sigma_{\rm HI}$ denoting the photoionization absorption cross-section for photons at
the hydrogen photoionization edge. This gives
\beqa
\Delta_{\rm ss} = 15 \left[\frac{\Gamma_{\rm HI}}{10^{-13} s^{-1}}\right]^{2/3} \left[\frac{1+z}{7}\right]^{-3} \left[\frac{T}{10^4 K}\right]^{0.133},
\label{eq:delta_ss}
\eeqa
for the case of isothermal gas (e.g. \cite{Furlanetto:2005xx}). The value adopted here for the photoionization rate, $\Gamma_{\rm HI} \sim 10^{-13} s^{-1}$, is comparable to the average photoionization
rate inferred from the $z \sim 5-6$ Ly-$\alpha$ forest \cite{Bolton:2007b,2008ApJ...688...85F}.

This simple estimate gives some sense for which regions of the IGM manage to self-shield and stay partly neutral during reionization, but recent 
simulation work has addressed this problem in greater detail. A key point is that the gas just at the self-shielding threshold of Eq. \ref{eq:delta_ss}
is still typically highly-ionized. It is then is a poor approximation to assume, for instance, that all gas with overdensity larger than $\Delta_{\rm ss}$ is highly neutral while
all gas below this density is completely optically thin. In order to study this, several works have performed radiative transfer calculations in a post-processing step
on top of evolved Smooth-Particle Hydrodynamics (SPH) simulations and quantified the trend of neutral fraction with density in partly shielded regions \cite{2011ApJ...737L..37A,McQuinn:2011aa,2013MNRAS.430.2427R}. These studies consider the attenuation of a uniform
UVB with a power-law or Haardt \& Madau \cite{Haardt:2001zf} spectrum by dense absorbers in the SPH simulation. Note that the results depend somewhat on the spectral shape and intensity of the ionizing radiation.
The study of Rahmati et al. (2013) \cite{2013MNRAS.430.2427R} includes a useful fitting formula to their simulation results, describing the attenuation of the (Haardt \& Madau 2001~\cite{Haardt:2001zf}) background radiation field by an overdense, partly-shielded region:
\beqa
\frac{\Gamma_{\rm att}}{\Gamma_{\rm UVB}} = 0.98 \times \left[1 + \left(\frac{\Delta}{\Delta_{\rm ss}}\right)^{1.64}\right]^{-2.28}  + 0.02 \times \left[1 + \frac{\Delta}{\Delta_{\rm ss}}\right]^{-0.84}.
\label{eq:ss_fit}
\eeqa
Here $\Gamma_{\rm UVB}$ is the intensity of the ultraviolet ``background'' radiation incident on the absorber of density $\Delta$, while $\Gamma_{\rm att}$ denotes the attenuated radiation
experienced by the absorber interior. The expression also involves the characteristic self-shielding overdensity, $\Delta_{\rm ss}$, of Eq. \ref{eq:delta_ss}. From the attenuated photoionization rate, $\Gamma_{\rm att}$,
one can determine the neutral fraction of the absorber according to the condition of photoionization equilibrium. This fitting formula then specifies the neutral fraction of dense clumps
given the gas overdensity (implicitly smoothed on the local Jeans scale), and the incident photoionization rate, $\Gamma_{\rm UVB}$. 

The above fitting formula can be used as the basis for a sub-grid model describing the photon sinks during reionization, as done in recent work by Sobacchi \& Mesinger (2014) \cite{Sobacchi:2014rua}. Note, however, that the above formula
still requires the local gas overdensity $\Delta$ as input, which requires resolving the Jeans scale -- since this is impossible for current large volume reionization simulations, further approximations are necessary.
In \cite{Sobacchi:2014rua}, the authors added small-scale structure to coarse simulation cells by drawing from the Miralda-Escud\'e et al. (2000) fitting formula for the gas density PDF \cite{MiraldaEscude:1998qs} (see \S \ref{sec:gas_pdf} below) . They assume
an approximate redshift evolution for the small scale structure in each cell, so that the density in each cell evolves in a sensible way with redshift. However, this approach ignores correlations between
the small-scale structure in neighboring cells and will not yield the correct small scale matter power spectrum. Nonetheless, it succeeds in capturing the main effect of dense clumps and their important 
impact on the process of reionization itself, as we further describe below.

\section{Spatial Fluctuations in the Ionization Fraction}
\label{sec:bubbles}

In addition to the global average ionization fraction, it is also interesting and useful to consider the spatial fluctuations in the ionized fraction at different stages of
the reionization process, i.e. to study ionization fluctuations for various values of $\avg{x_i}$ and redshift. The spatial fluctuations in the ionization fraction are mostly sensitive to different aspects
of the reionization model than the average history is, and also impact observational probes of reionization differently than the mean history, and so it is useful to consider the fluctuations separately.
Indeed, the {\em inhomogeneities} in the reionization process are important for a broad range of reionization observables, especially studies of the redshifted 21 cm line \cite{Furlanetto:2004ha,Mellema:2006pd,Lidz:2007az,2011MNRAS.411..955M}, narrow band surveys for Ly-$\alpha$ emitters \cite{McQuinn:2007dy}, small-scale CMB anisotropies through the patchy kSZ effect \cite{Zahn12,Mesinger12}, Ly-$\alpha$ forest absorption spectra \cite{Lidz:2007mz,Mesinger:2009mv,Malloy:2014tba}, and observations of gamma-ray burst optical afterglows \cite{McQuinn:2007gm,Mesinger:2007kd}, among others. 
We should keep in mind, however, that the mean history is {\em not completely decoupled} from the fluctuations in the ionized fraction: the clumping factors and source-absorber correlations impact
the mean history and depend on
the spatial variations in the ionization field.

As discussed earlier, provided star forming galaxies are the ionizing sources, the ionization state of the IGM will resemble a two-phase medium with a mix of mostly ionized bubbles and intervening
highly neutral regions. The study of spatial fluctuations in the ionization fraction then amounts mostly to examining the size distribution of the ionized bubbles as a function of $\avg{x_i}$ and $z$.
The bubbles of ionized gas will, however, contain some dense regions that manage to self-shield and remain partly or largely neutral, as described in the previous section. While these dense
clumps should fill only a small fraction of the volume, they nevertheless play an important role in determining the size of the ionized bubbles because they consume ionizing photons and thereby slow
the growth of the ionized regions. 

The spatial variations in the ionization fraction mostly reflect spatial variations in the timing of reionization across the universe, with galaxy formation starting earlier, and reionization completing more quickly,
in some regions of the universe \cite{Barkana:2003qk}. This notion can be used to develop some intuition for how the size distribution of the ionized regions depends on the underlying reionization model \cite{Furlanetto:2004nh}.
In particular, dark matter halos -- and by extension galaxies -- form first in large scale overdensities: halos collapse first in the rare high-density peaks of the density distribution. The same overdense
regions will also generally contain more {\em sinks} of ionizing photons than typical regions: provided, however, that the sources of ionizing photons are more biased tracers of the density distribution than
the sinks,
the large scale overdense regions should reionize first. The sinks of ionizing photons will, however, slow the growth of the ionized regions around the large scale overdensities; this then allows the
ionized regions forming around galaxies that turn on somewhat later to partly catch up in size.
This intuition suggests that the size distribution of the ionized regions should depend on the bias of the
sources and the absorbers, i.e. on their spatial distribution and clustering. In addition, this reasoning implies that large-scale overdense regions should reionize first, while on small-scales overdense clumps should recombine rapidly and be ionized last. Hence whether
reionization is ``inside-out'' (overdense regions close to sources ionize first), or ``outside-in'' (overdense regions ionize last) is partly a question of scale.

Many of these qualitative features can be understood more quantitatively in the excursion set model of reionization \cite{Furlanetto:2004nh}. We will not review this model in detail here, and instead only briefly summarize the approach. 
In the simplest variant of this model, one supposes that
each halo above some minimum mass $M_{\rm min}$ is capable of ionizing a mass of intergalactic hydrogen atoms that is proportional to the halo mass. A region of large-scale overdensity, $\delta_M$ -- when the region is smoothed on mass scale $M$ -- can then be
ionized if a sufficient fraction of the mass in the region has collapsed into halos above $M_{\rm min}$. In other words, for a region to be ionized the conditional collapse fraction, $f_{\rm coll} (> M_{\rm min} | \delta_M, M)$,
must exceed some critical value.  The statistical properties of the ionized regions are then determined by considering the probability distribution that the (linear) density field exceeds this critical value for different smoothing scales. The semi-numeric technique developed in \cite{Zahn:2006sg, Mesinger:2007pd} and other work essentially implements this criterion by smoothing simulated 
three-dimensional realizations of the initial density field, or by applying a related criterion to the evolved dark matter halo distribution. For more details here the reader may refer to the original papers and \cite{2013fgu..book.....L}.

\begin{figure}[b]
\sidecaption
\includegraphics[scale=0.65]{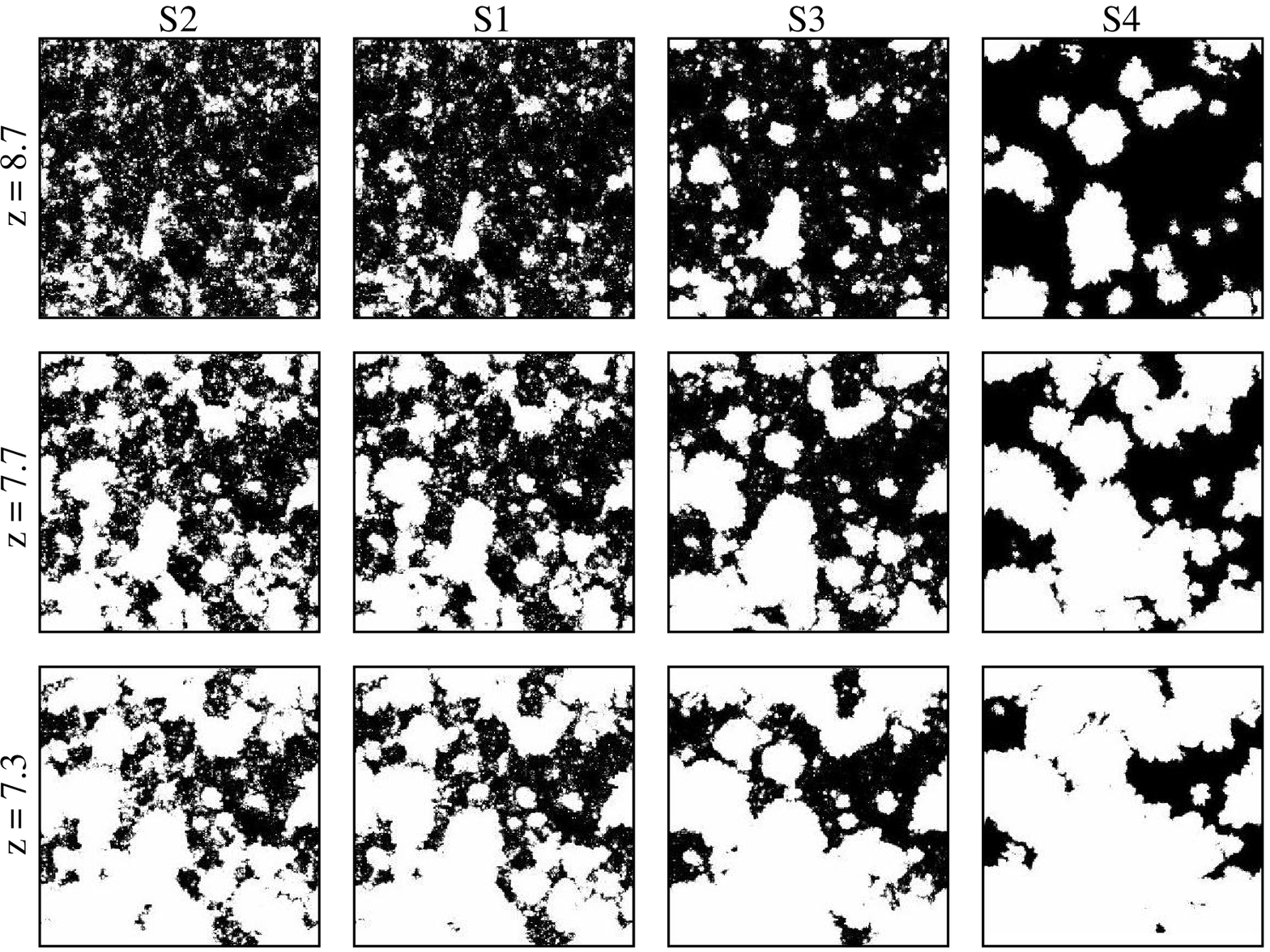}
\caption{Dependence of bubble sizes on ionizing source properties. The panels show slices through the ionization field from radiative transfer simulations of reionization with varying prescriptions
for the properties of the ionizing sources. Each panel is $0.25$ Mpc/$h$ thick, and spans $65.6$ co-moving Mpc/$h$ on a side. The white regions show highly ionized gas in the simulation while the
dark regions are neutral.  
The columns, moving from left to right, show different models for the ionizing sources: the left-most column assumes
that the rate of ionizing photons emitted scales with host halo mass as $\dot{N} \propto M^{1/3}$, the left-center assumes $\dot{N} \propto M$, the right-center takes $\dot{N} \propto M^{5/3}$, and
the right-most has $\dot{N} \propto M$, but increases the minimum host halo mass from $M_{\rm min} = 10^8 M_\odot$ to $M_{\rm min} = 4 \times 10^{10} M_\odot$. 
Moving from top to bottom, we show slices from various stages of reionization with volume-averaged ionized fractions of
$\avg{x_i} \approx 0.2, 0.5,$ and $0.7$, respectively.  The bubble sizes, at a given stage of reionization, are larger in the case where reionization is mainly driven by more massive, highly biased 
sources. From \cite{McQuinn:2006et}.}
\label{fig:bubble_sizes}
\end{figure}

The spatial variations in the ionization field may also be studied using radiative transfer simulations. Encouragingly, a detailed comparison study \cite{2011MNRAS.414..727Z} between two different radiative transfer simulations \cite{McQuinn:2007dy,Trac:2006vr} and semi-numeric calculations, demonstrates remarkably good agreement between the two simulations, {\em and} with the semi-numeric calculations.
While this study and the earlier work of \cite{Zahn:2006sg} help to validate the semi-numeric approach as a useful tool, and demonstrate that many features of the radiative transfer
calculations can be understood in a simple manner, we should keep in mind that fairly simplistic models for the ionizing sources were assumed in both the radiative transfer simulations and the semi-numeric models in these comparisons. In addition, these calculations did not capture dense photon sinks and so further work is required here.

As one example, Fig. \ref{fig:bubble_sizes} provides a visual illustration of how the sizes of the ionized regions depend on the bias of the ionizing sources, at various stages of the reionization process. The different panels show 
slices through a radiative transfer simulation of reionization from \cite{McQuinn:2006et}. In each model, the radiative transfer calculations were performed in a post-processing step on top of an evolved N-body (gravity only) simulation. The ionizing sources are placed in simulated dark matter halos, with various prescriptions for connecting the ionizing luminosity with the host halo mass. In particular, as one
moves from the left-most to the right-most panels in Fig. \ref{fig:bubble_sizes}, the ionizing luminosity is dominated by sources in progressively more massive host dark matter halos. In each row,
the normalization of the ionizing luminosity-host halo mass relation is set  to produce the same volume-averaged ionization fraction ($\avg{x_i}$), with $\avg{x_i} \approx 0.2, 0.5$,
and $0.7$ (from top to bottom), so the different rows correspond to early, middle, and later stages of the reionization process.  The figure illustrates that the ionized regions are larger, at each stage
of the reionization process, in models in which the ionizing sources reside mostly in progressively more massive, and hence more {\em biased}, host halos. In the right-most panel, however, the ionizing sources are in very rare, yet individually luminous, host halos ($M \geq 4 \times 10^{10} M_\odot$); in this case, the ionized regions are large enough (for the $\avg{x_i}$ shown) to enclose underdense and overdense regions alike and the correlation between the ionization and density fields is correspondingly smaller (e.g. \cite{Lidz:2006vj}). Put differently, for sufficiently rare sources Poisson fluctuations in the source abundance dominate over density-sourced variations (on the scale of the ionized bubbles) and this removes the trend for large-scale overdensities to reionize first.
This final Poisson-dominated case, however, seems unlikely given that measurements of the UV galaxy luminosity function suggest that numerous individually faint
galaxies reionized the universe, as discussed in \S \ref{sec:sources_further}.

\begin{figure}[b]
\sidecaption
\includegraphics[width=3.5cm]{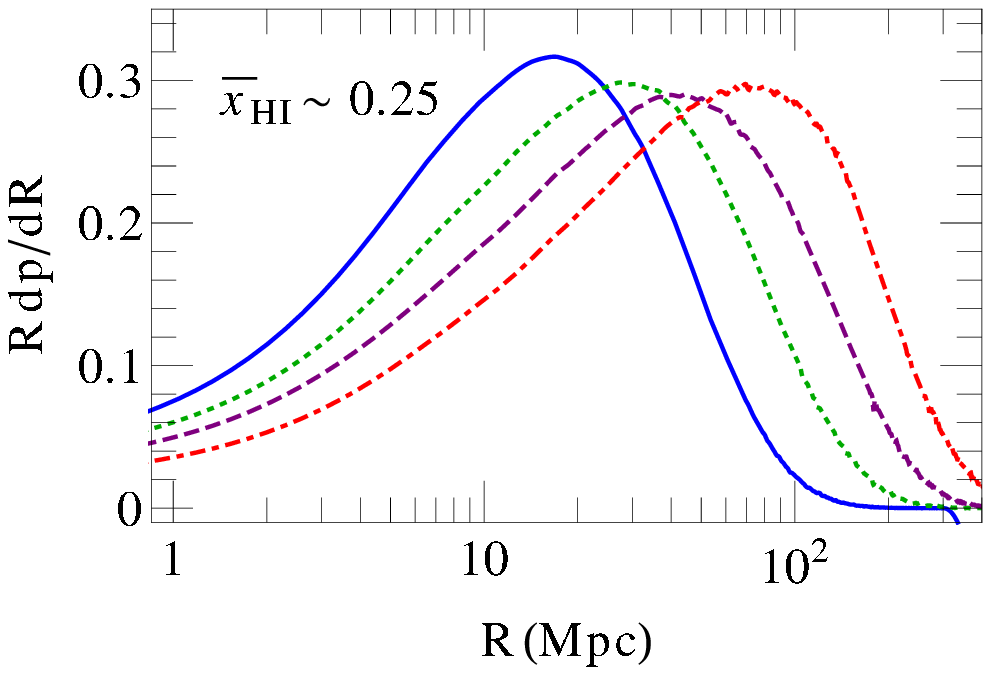}
\includegraphics[width=3.5cm]{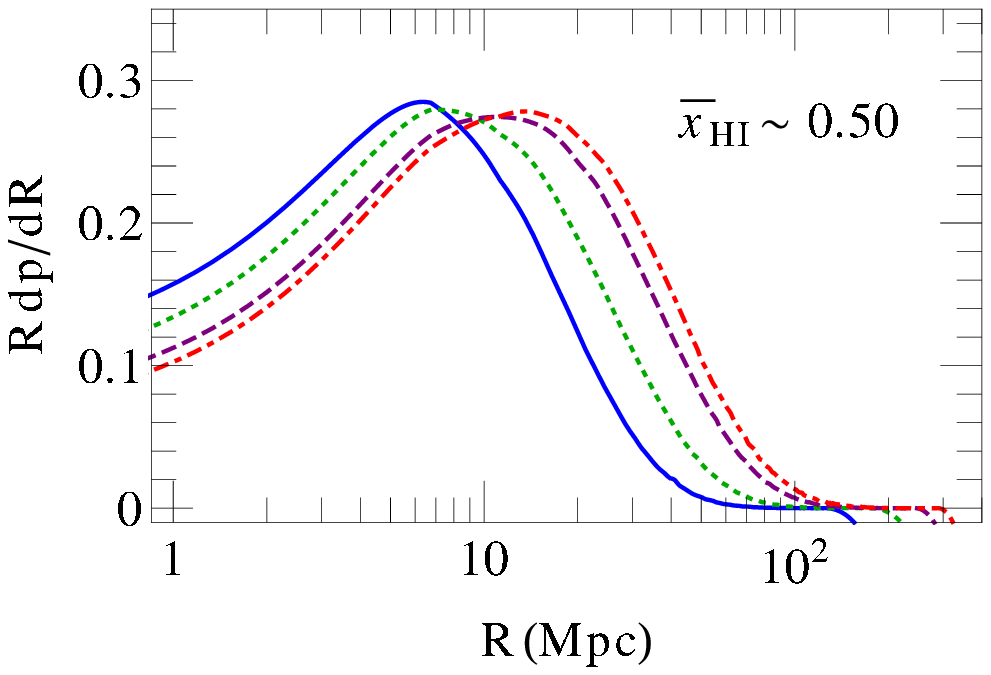}
\includegraphics[width=3.5cm]{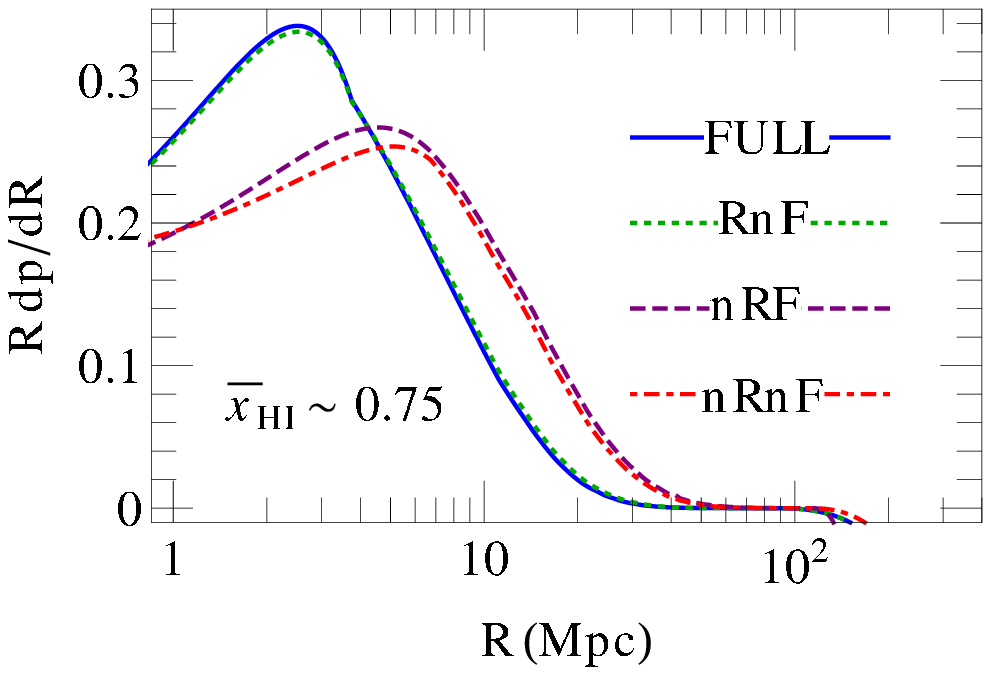}
\includegraphics[width=3.5cm]{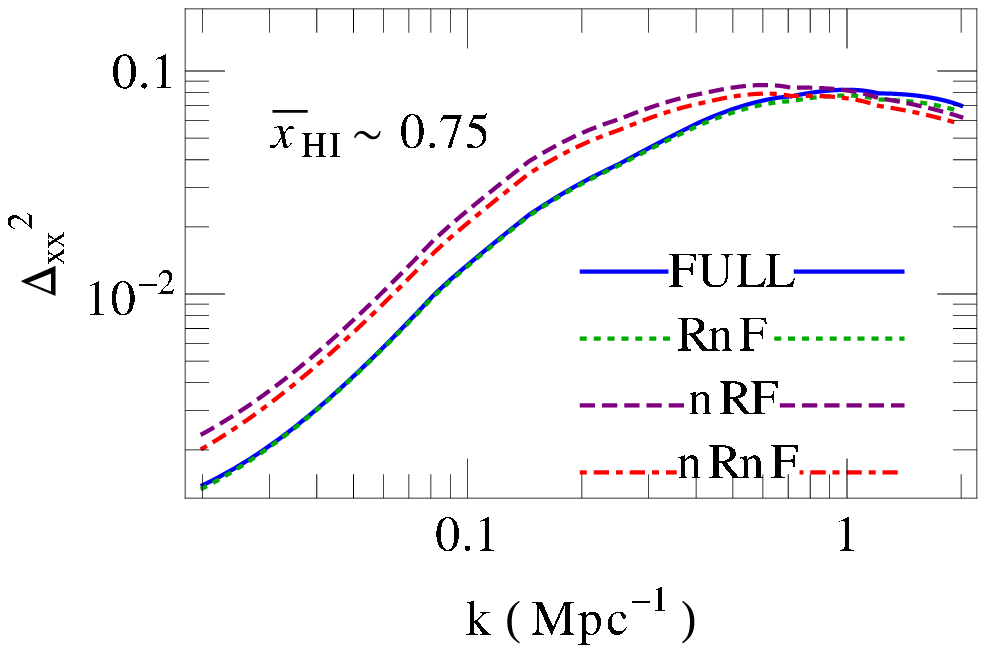}
\includegraphics[width=3.5cm]{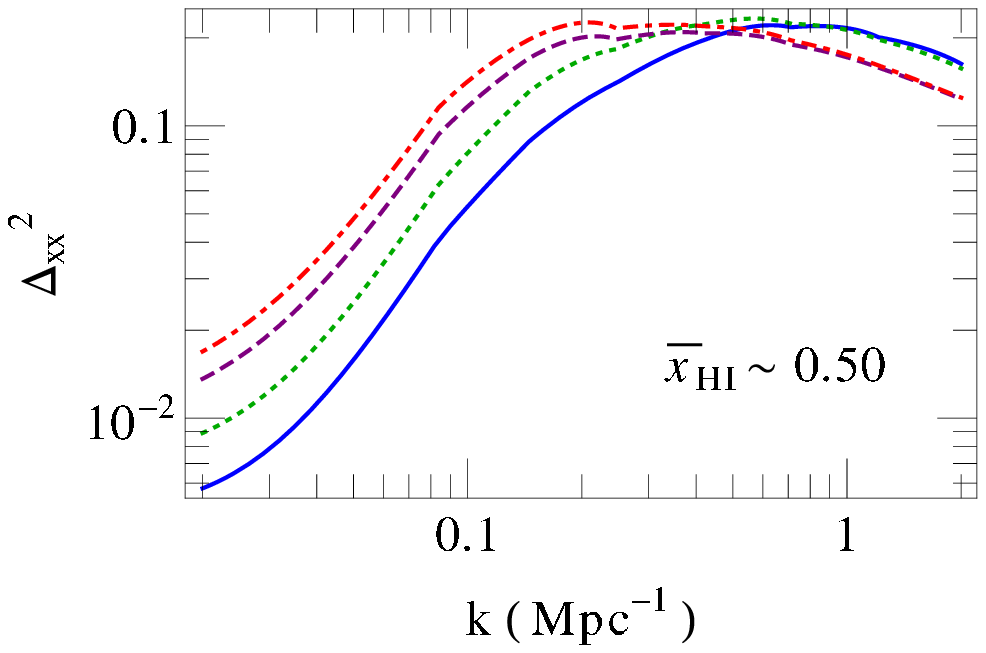}
\includegraphics[width=3.5cm]{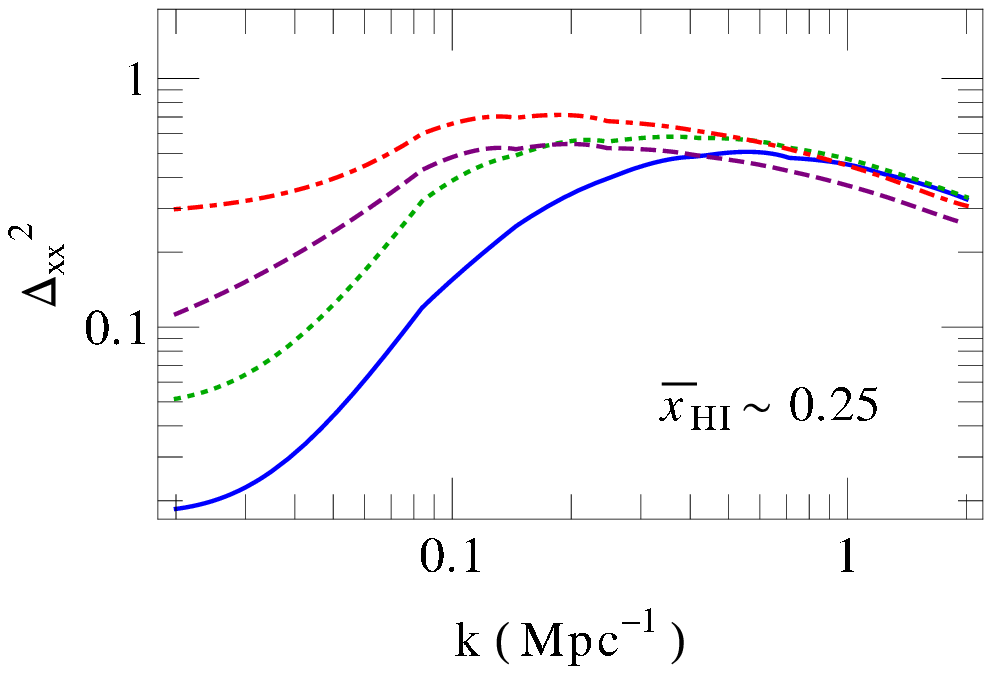}
\caption{Model bubble size distributions and ionization power spectra. The upper row shows the size distribution of ionized regions at different
stages of the reionization process: from left to right, the volume-averaged neutral fraction is $\avg{x_{\rm HI}} \approx 0.25, 0.5,$ and $0.75$, respectively. 
The blue solid curve includes a model for the impact of recombinations in dense, self-shielded clumps, which act to limit bubble growth. This case also includes a sub-grid
model for the impact of photoionization feedback, which here boosts the minimum host halo mass in ionized regions depending on the intensity of the local radiation field and the length
of time the host halo has been exposed to this radiation. The green dotted line includes the impact of recombinations in dense clumps, but ignores the effects
of photoionization feedback. The purple dashed line ignores recombinations, but includes  photoionization feedback. Finally, the red dot-dashed line ignores both recombinations
and photoionization feedback. The bottom row shows the power spectrum in the same set of models at each of $\avg{x_{\rm HI}} \approx 0.75, 0.5,$ and $0.25$. From \cite{Sobacchi:2014rua}. }
\label{fig:sobacchi_bubbles}
\end{figure}

As we anticipated earlier, the next important determinant of the bubble sizes are the properties and the spatial distribution of the sinks of ionizing photons. 
Note that the impact of the sinks of ionizing photons 
is inter-connected with how quickly the radiation from the ionizing sources is able to build up \cite{Choudhury:2008aw}. The sinks of ionizing photons have less time to act and play a less important role if the ionizing sources
are especially luminous and turn on quickly. If, however, the sources turn on more gradually -- and are less luminous -- then this allows more regions in the IGM to self-shield and the dense clumps
are able to consume significant numbers of ionizing photons. As discussed earlier, present evidence prefers a low post-reionization emissivity, favoring a gradual reionization process and an
important role for the ionizing sinks in regulating bubble sizes.

Fig. \ref{fig:sobacchi_bubbles} (from \cite{Sobacchi:2014rua}) illustrates how the sinks of ionizing photons may limit the growth of ionized regions. These authors use a sub-grid model, based on the discussion of the previous
section, to incorporate dense photon sinks into a semi-numerical reionization simulation.  These models are in part calibrated to reproduce the mean free path to ionizing photons after reionization \cite{2010ApJ...721.1448S,Worseck:2014fya}.
The top row of Fig. \ref{fig:sobacchi_bubbles} compares the probability distribution function (PDF) of the ionized bubble sizes for simulations that include photon sinks with calculations that do not
include these sinks. Some of the models in the figure also incorporate the impact of photoionization feedback, which acts to boost the minimum host halo mass in the ionized regions.
The difference between the green dotted (including recombinations but no photoionization feedback) and the red dot-dashed models (ignoring both recombinations and photoionization
feedback), shows the impact of the dense photon sinks on the bubble sizes. The photon sinks reduce the size of the ionized regions by a factor of $\sim 2-3$,  in comparison with the model that neglects
these sinks.  Photoionization feedback further reduces the bubble sizes somewhat.

The presence of dense photon sinks has two separate, but related, effects on the reionization process. First, for bubble growth to continue, the {\em rate} at which ionizing photons are produced by
the sources interior to the bubble (or supplied by neighboring sources) must exceed the rate at which atoms in the interior recombine. If the ionized bubbles grow to the point that they enclose
a sufficient number of dense clumps, the high rate of recombinations in the interior clumps will match the rate at which the sources produce the ionizing photons, and the bubble growth will halt (e.g. \cite{Furlanetto:2005xx}). This recombination-limited
growth is analogous to the familiar case of a Stromgren sphere forming around a massive star in the ISM of a galaxy, except that here the ionized regions generally grow under the collective influence of thousands of galaxies. The second effect is that for an ionized bubble to grow, the cumulative number of ionizing photons received within the bubble must exceed the number of interior atoms plus the number of
interior recombinations. In other words, the first criterion places a requirement on the instantaneous rate of production of ionizing photons, while the second one demands some cumulative output
of ionizing photons. The study of \cite{Sobacchi:2014rua} finds that this latter requirement is more important, at least for the model of the ionizing sources and sinks considered in that work.

Another useful statistic for quantifying the spatial fluctuations in the ionization, or neutral fraction, field is the power spectrum. 
The bottom row of Fig. \ref{fig:sobacchi_bubbles} shows the power spectrum of the fluctuations in the neutral fraction from the Sobacchi \& Mesinger (2014) models \cite{Sobacchi:2014rua}. Defining the fractional fluctuation in the neutral fraction as $\delta_x = (x_{\rm HI} - \avg{x_{\rm HI}})/\avg{x_{\rm HI}}$, this row shows $\Delta^2_{xx}(k) = k^3 P_{xx}(k)/2 \pi^2$, i.e. the contribution to the variance of the neutral fraction field per logarithmic interval in $k$. The photon
sinks reduce the large scale power in the neutral fraction field in each panel, by reducing the size of the ionized regions at each stage of reionization. As the ionized regions grow, they imprint
large scale fluctuations in the neutral fraction field -- this boosts the amplitude of the $\Delta^2_{xx}(k)$ power spectrum on large scales. This quantity becomes flat in comparison to the density power
spectrum on scales in which there are a sufficient number of ionized regions. The photon sinks act to reduce the scale of the largest ionized regions, and so the power spectrum doesn't become as flat
at small $k$ as it does without the dense photon sinks. This effect can be seen most prominently by comparing the models with and without dense photon sinks at $\avg{x_{\rm HI}} \sim 0.75$ and $k \lesssim 0.1$ Mpc$^{-1}$ in the bottom right panel of Fig. \ref{fig:sobacchi_bubbles}.  Upcoming redshifted 21 cm surveys will measure the power spectrum of 21 cm brightness temperature fluctuations (closely related
to the neutral fraction fluctuation power spectrum shown here) near these scales (e.g. \cite{Lidz:2007az}), and so the impact of photon sinks has important implications for these measurements. The reduced bubble sizes, and reduced
large-scale fluctuation power, in the models with dense photon sinks generally imply that the redshifted 21 cm power spectrum during reionization will be harder to detect than previously thought.
 
\section{Spatial Fluctuations in the UV Radiation Background}
\label{sec:uvb_fluc}

In addition to the order unity spatial fluctuations in the ionization fraction owing to the ionized bubbles, there will also be fluctuations in the ionized fraction within the bubbles
themselves. These fluctuations result because overdense regions recombine more rapidly than typical regions, while sufficiently overdense clumps can self-shield from the
UV radiation background, and also because the intensity and spectrum of the UV radiation varies spatially. Spatial fluctuations in the temperature (see \S \ref{sec:temp_igm}) should also play  a role here because the recombination rate in the gas is temperature dependent. Here we focus on the spatial fluctuations in the UV radiation field: these impact both the location and properties of the dense photon sinks, as well as the
(varying) residual neutral fraction within the ionized regions. These properties then influence the growth of the ionized bubbles and the reionization process itself, as well as the mean free path to ionizing photons. Note
also that the fluctuations in the UV radiation field are determined in part by the properties of the dense clumps and the mean free path, which are in turn influenced by the surrounding UV radiation, and
so one needs to treat the sinks and sources of ionizing radiation self-consistently. In addition to influencing the overall process of reionization, radiation fluctuations impact the 
residual neutral fractions within the ionized regions, and may hence also be directly relevant for the interpretation of high redshift Ly-$\alpha$ forest spectra and narrow-band Ly-$\alpha$ emitter surveys. 
For example, the high optical depth in the Ly-$\alpha$ line at $z \gtrsim 5$ makes these surveys sensitive
to whether the residual neutral fraction within ionized regions is typically more like $x_{\rm HI} \sim 10^{-5}$ or instead $x_{\rm HI} \sim 10^{-4}$ \cite{1965ApJ...142.1633G}, which will be influenced
by the average UV radiation field and its spatial fluctuations.

It is instructive, once again, to first consider the case of the post-reionization IGM where we presently have more empirical guidance.
At redshifts of order $z \sim 3$, the mean free path to ionizing photons is quite large, on the order of a few hundred co-moinvg Mpc. For example, the preferred value from a recent measurement
of the mean free path to ionizing photons for photons at the hydrogen photoionization edge frequency gives $\lambda=220$ co-moving Mpc/$h$ at $z=3.6$ \cite{2009ApJ...705L.113P}. Each location in the IGM then typically sees the combined radiation from many, many sources: in this case, the fluctuations in the UV radiation field are quite gentle. As a result, it makes
sense to consider the UV radiation as a (nearly) uniform  {\em background}, and simulations adopting a uniform radiation field generally match the main
properties of the $z \sim 2-4$ Ly-$\alpha$ forest fairly well \cite{Meiksin:2003qb,Croft:2003qn,McDonald:2004xp}. However, this assumption will inevitably break down close to and during reionization, when the mean-free
path becomes short and also {\em spatially variable}. 

Let us start with a rough estimate for the strength of these fluctuations (e.g. \cite{Zuo92,Zuo93,2009MNRAS.400.1461M}). Specifically, let us consider the Poisson fluctuations and the cosmological fluctuations (owing to variations in the underlying density field) in the number of sources contained within a sphere
of radius equal to the mean free path. If the average number density of sources is $\avg{n}$, then the average number of sources within
the mean-free path sphere is $N_s = \avg{n} 4 \pi \lambda^3/3$, where we adopt co-moving units for the mean-free path, $\lambda$, and the average source
density, $\avg{n}$. As above, $\lambda$ is the mean-free path to photons at the hydrogen ionization edge and likewise $J$ is shorthand for the specific intensity at the ionization edge.
The fractional fluctuations in the radiation background owing to Poisson fluctuations  scale as $\sigma_J/\avg{J} \sim 1/\sqrt{N_s}$, while
the spatial variations from large scale structure give $\sigma_J/\avg{J} \sim \avg{b} \sigma_{\delta}(\lambda)$. Here $\avg{b}$ is the typical (luminosity-weighted) bias of the ionizing
sources, and $\sigma_\delta(\lambda)$ is the square root of the matter density variance, smoothed on the scale of the mean free path. It is evident that these
variations will be small when the mean free path is sufficiently large. In addition to these variations, regions of the IGM that happen to lie very close to a bright
source will see an enhanced radiation background owing to the nearby source. However, this ``proximity scale'' is generally small unless the local source is bright, while sufficiently bright sources are rare. 
The proximity scale is $r_p \sim \left[L_s/(4 \pi \avg{n L} \lambda) \right]^{1/2}$ where $L_s$ is the luminosity of the local source and $\avg{n L}$ is the emissivity of the
background sources. Suppose that the local source has a typical luminosity such that $L_s/\avg{n L} = 1/\avg{n}$. Then let $r_0$ be the radius of the sphere that contains one source on average, so
that $4 \pi r_0^3 \avg{n}/3 = 1$. It then follows that $r_p/r_0 =  \left[r_0/(3 \lambda)\right]^{1/2}$. Hence the proximity scale is typically small compared to the average separation between the ionizing
sources, provided the mean free path is large compared to the average source separation, i.e., $r_p << r_0$ provided $\lambda >> r_0$. After reionization, the mean
free path must be large compared to the average source separation (essentially by definition). Even during reionization, the ionized regions typically grow under the collective influence of many ionizing sources, and
so the typical proximity scale should still be fairly small relative to the average source separation.

We can make these estimates more explicit as follows.  Let us consider the specific intensity, $J_\nu$, (with units of ergs s$^{-1}$ cm$^{-2}$ Hz$^{-1}$ str$^{-1}$) of ionizing radiation incident on gas elements in the IGM and the spatial fluctuations in this quantity. Here we make four main approximations which have been adopted in most previous work,  although the approximations involved are not always clearly stated in the literature. First, let us take the case that the mean free path to ionizing photons is small compared
to the Hubble radius at the redshift of interest, so we can neglect the redshifting of photons. This should be a good approximation at $z >> 2$ \cite{1999ApJ...514..648M}.  Second, we ignore spatial variations in the attenuation length, and assume that
there is a single mean free path to ionizing photons across the entire universe. This assumption must actually be a poor approximation during reionization: for instance, during some phases of the EoR, the sizes of the ionized regions mostly set the mean free path and the ionized regions certainly have a broad range of sizes.
Even when the mean free path is mostly set by dense clumps/Lyman limit systems, there
will be significant variations in their abundance. For example, in regions with enhanced levels of ionizing radiation, the ionizing radiation will penetrate more deeply into the dense clumps and this will
enhance the mean free path locally. Third and somewhat related, we ignore spatial correlations between the sources and the absorbers. This assumption again appears of dubious validity near the EoR: the sinks of ionizing photons
will generally be spatially correlated with the sources  and so overdense regions should typically contain {\em more absorbers} as well as more sources, although the enhanced radiation field
in an overdense region will counteract this \cite{Crociani11}. Fourth, we ignore spatial variations in the spectral shape of the ionizing radiation.  In general, one would expect there to be spatial variations in
the amount of spectral hardening as radiation arrives at a given location from a variety of different pathways. Some radiation will have crossed through filaments and dense clumps that remove all but
the highest energy photons, while other radiation will travel mostly along underdensities and suffer less hardening. These effects presumably average down and are less important when the mean free path
is long. In this case, the ionizing radiation incident on a typical point in the IGM comes from many different sources and the ionizing photons generally arrive at this point after propagating across long distances and sampling a range of different environments.

Adopting these approximations, the spatially-averaged specific intensity is:
\beqa
\avg{J_\nu} = (1+z)^2 \frac{\avg{\epsilon_\nu} \lambda_\nu}{4 \pi}.
\label{eq:avg_jnu}
\eeqa
Note that the factor of $(1+z)^2$ enters because here we are taking $\lambda_\nu$ to be the co-moving mean-free path and the emissivity in the above expression, $\avg{\epsilon}$, denotes the emissivity per co-moving volume.
As discussed previously (\S \ref{sec:sources_further}), the mean intensity of the radiation at $z \leq 6$ can be constrained from the average transmission through the Ly-$\alpha$ forest.

To reiterate briefly, the formula above ignores spatial correlations between the sources and absorbers and fluctuations in the attenuation length. Correlations (or anti-correlations) between the sources and the absorbers would imply that the average of the product of the emissivity and attenuation length is {\em not} the same as the product of the average of these two quantities, contrary to what is assumed here. We further assume
that the attenuation in the specific intensity around each source is characterized by $J^S_\nu(r) \propto (L^S_\nu/4 \pi r^2) {\rm exp}(-r/\lambda_\nu)$, where the superscript $S$ denotes the contribution
from a particular source. The spatial fluctuations in $J_\nu$ at each particular frequency -- we suppress the notation $\nu$ below to make the notation compact -- can then be characterized by a power spectrum:
\beqa
\frac{P_J(k)}{\avg{J}^2} = \left[\frac{{\rm arctan}(k \lambda)}{k \lambda} \right]^2 \left[\avg{b}^2 P_{\rm lin}(k) + \frac{\avg{L^2}}{\avg{L}^2} \right].
\label{eq:pjofk}
\eeqa
Here $\avg{b}$ is an average luminosity-weighted bias of the ionizing sources, and the second term results from Poisson fluctuations in the source abundance. The modeling here -- and the two separate contributions to the radiation fluctuations -- are analogous to the
``halo model'' for the matter power spectrum \cite{Cooray:2002dia}, except here the mass profile around each source is replaced by a luminosity profile \cite{2009MNRAS.400.1461M}. 
The term in front is the ``window'' function
that results from the Fourier transform of $J^S(r) \propto {\rm exp}(-r/\lambda)/r^2$.  Note that in the limit $k \lambda << 1$ the window function $|W(k)|^2 = |{\rm arctan}(k \lambda)/(k \lambda)|^2 \rightarrow 1$, while on scales small compared to the mean free path, $|W(k)|^2 \propto 1/k^2$, and the power spectrum turns over.  According to this expression, the variance per logarithmic interval in $k$, 
$\Delta^2_J(k) = k^3 P_J(k)/(2 \pi^2)$, formally diverges at high $k$. However this is an artifact of assuming the flux around a sources scales as $J^S(r) \propto {\rm exp}(-r/\lambda)/r^2$ for
arbitrarily small $r$. In practice, the proximity scale where local sources dominate should be a small fraction of the mean separation between sources as discussed above.  Moreover, the source flux will
also not in reality vary as $1/r^2$ to arbitrarily small $r$.

\begin{figure}[b]
\sidecaption
\includegraphics[width=11cm]{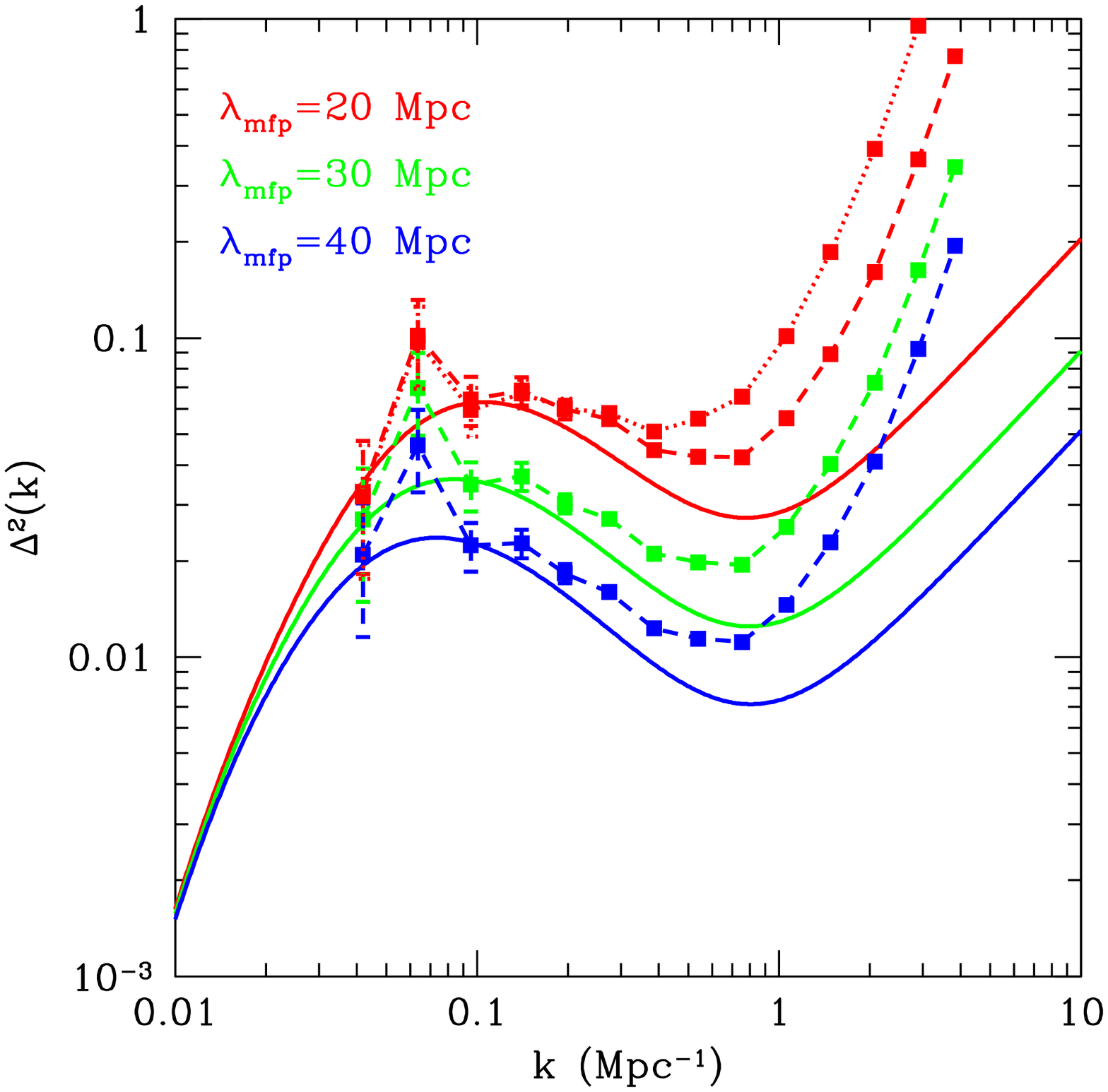}
\caption{Power spectrum of UVB fluctuations at $z=6$. The solid lines show models of the form of Eq. \ref{eq:pjofk} for $\Delta^2_J(k)$. In the models shown, $10\%$ of randomly selected halos more massive
than $M_{\rm min} = 1.6 \times 10^8 M_\odot$ host ionizing sources with an ionizing luminosity proportional to the host halo mass. The different colored lines make different assumptions about
the mean free path to ionizing photons. The dashed lines and squares show results from semi-numeric simulations at $z=6$, while the dotted lines and corresponding squares are for $z=5$. From \cite{2009MNRAS.400.1461M}.}
\label{fig:uvfluc}
\end{figure}

The results of calculations along these lines are shown in Fig. \ref{fig:uvfluc} at $z=6$ for various plausible values of the mean free path to ionizing photons (from \cite{2009MNRAS.400.1461M}).
On large scales, the clustering contribution to the UV fluctuations dominates while on small scales Poisson fluctuations in the abundance of the sources dominate. The amplitude of fluctuations drops
off on scales smaller than the mean free path to ionizing photons until on still smaller scales Poisson fluctuations lead to a $\Delta^2_J \propto k$ behavior at high wavenumbers.  The analytic calculations
(following Eq. \ref{eq:pjofk} above) match the semi-numeric simulations on large scales, but there is somewhat more power in the semi-numeric models on small spatial scales; this excess power likely results from non-linearities in the clustering that are not incorporated in the analytic model.  On large spatial scales, the peak in the power spectrum
corresponds to spatial fluctuations of amplitude $\sqrt{\Delta^2_J(k)} \sim 25 \%$ for $\lambda = 20$ Mpc. Although these are fairly strong fluctuations, the direct impact on the Ly-$\alpha$ forest
appears to be relatively small \cite{Meiksin:2003qb,2009MNRAS.400.1461M}. An important caveat here, however, is that these models have thus far ignored spatial variations in the attenuation length.
Indeed, recent Ly-$\alpha$ observations indicate a significant sightline-to-sightline scatter in the average transmission through the forest \cite{Becker:2014oga}, above that expected from density fluctuations alone \cite{Lidz:2005gn}. 
Further study here may provide additional insight into the precise impact of these fluctuations on the Ly-$\alpha$ forest and other observables, and regarding their role in the reionization process itself.

\section{The Temperature of the IGM}
\label{sec:temp_igm}

The temperature of the gas in the IGM is important because it is a basic property of the IGM, impacting the rate at which 
gas recombines and cools, as well as its ability to fall into the shallow potential wells of low mass dark matter halos (and
hence the minimum host halo mass of a galaxy, \S \ref{sec:sources}). 
It is also important because the 
low density gas retains some memory of when and how it was reionized \cite{1994MNRAS.266..343M,Hui:1997dp}. This ``memory'' effect is a consequence of the long cooling time for (most of) the IGM gas, and implies that measurements of the temperature of the IGM
after reionization can be used to constrain reionization itself \cite{Theuns:2002yc,Hui:2003hn}.  Here we summarize the main features of models of the IGM
temperature after reionization.

Let us consider the temperature evolution of a gas element after reionization. First we would like to quantify the energy input by photoheating
during reionization. Suppose the gas element has a volume $V$ and a total hydrogen number density $n_H$ and a helium number
density of $n_{\rm He}$.  As an ionization front sweeps across the gas element during reionization, we assume that the hydrogen in
the gas element becomes highly ionized and the helium mostly singly ionized. We further assume that the specific intensity of ultraviolet radiation produced by 
the ionizing sources is well approximated by a power-law between the hydrogen ionization edge (at frequency $\nu = \nu_{\rm HI}$) and the threshold frequency
to doubly ionize helium (at $\nu = 4 \nu_{\rm HI}$), but that the sources emit negligibly few photons capable of doubly-ionizing helium. In other words,
we take $J(\nu) \propto \nu^{-\alpha}$ for
$\nu_{\rm HI} \leq \nu \leq 4 \nu_{\rm HI}$, and assume the spectrum is truncated at higher frequencies.
Finally, let us take the gas element to be initially at low temperature so that its thermal energy increases significantly
after it is reionized, and consider the limit that the gas is completely optically thick (below $4 \nu_{\rm HI}$) before reionization.  
In this case, the
average energy input by photoionization is (e.g. \cite{Abel:1999vj,Furlanetto:2009kr}):
\beq
\avg{\Delta E} = n_H V \frac{\int_{\nu_{\rm HI}}^{4 \nu_{\rm HI}} \frac{d\nu}{h\nu} J(\nu) (h\nu - h\nu_{\rm HI})}{\int_{\nu_{\rm HI}}^{4 \nu_{\rm HI}} \frac{d\nu}{h\nu} J(\nu)}.
\label{eq:del_e}
\eeq
On a short time scale, the photoelectrons will share their energy with the surrounding gas so that $\avg{\Delta E} \approx
3 (n_H + n_{He}) V k_b T_{\rm reion}$, where the number densities here assume that hydrogen is highly ionized and helium mostly singly ionized: accounting
for free electrons, the number of free particles in the gas element is $2 \times (n_H + n_{He}) V$ and the average thermal energy per free particle in the gas is $3 k_b T/2$.
Note that Eq. \ref{eq:del_e} does not include a weighting by the photoionization absorption cross section, because we adopt the optically thick limit in which all of the
incident radiation between one and four Rydbergs is absorbed by the gas \cite{Abel:1999vj}.
The temperature of the recently photo-ionized
gas at reionization is 
\beq
T_{\rm reion} \approx \frac{n_H}{3 (n_H + n_{He})} h \nu_{\rm HI} \frac{1 - 4^{-\alpha} - 3 \alpha~ 4^{-\alpha}}{\left(\alpha-1\right)\left(1-4^{-\alpha}\right)}.
\label{eq:treion}
\eeq
For a plausible star-forming galaxy spectrum with $\alpha=2$, this estimate gives $T_{\rm reion} \sim 3 \times 10^4$ K. (See e.g. \cite{Furlanetto:2009kr} for further discussion).

Next we want to consider the {\em evolution} of the temperature of each gas element after reaching a temperature $\sim T_{\rm reion}$ during
reionization. After reionization, the dominant cooling processes are adiabatic cooling from the expansion of low density gas elements and
Compton cooling as CMB photons scatter off of free electrons and extract thermal energy from the gas. The UV radiation from star-forming galaxies
keeps the gas highly ionized after reionization, but there is still some residual photoionization heating as ionization equilibrium is maintained.
The temperature evolution of a gas element can be determined by applying the first law of thermodynamics, $dE=-P dV + dQ$, to a gas
element of fixed total mass. The thermal energy of the gas is $E=3 k_b T \rho V/(2 \mu m_p)$, with $\mu$ denoting the mean mass per particle in the gas
in units of the proton mass, while the pressure of the gas is given by the ideal gas law, $P = \rho k_B T/(\mu m_p)$. The impact of photoionization
heating and cooling processes is to inject/remove heat, $dQ$, into the gas as described by heating and cooling functions, $dQ = ({\mathcal H} - \Lambda) V dt$.
Here $dt$ is a time increment, and so with the definition adopted here, ${\mathcal H}$ and $\Lambda$ have units of energy per volume per time.
The resulting thermal evolution equation is (e.g. \cite{Hui:1997dp}):
\begin{align}
\frac{dT}{dt} =& -2 H T + \frac{2 T}{3 (1+\delta)}\frac{d\delta}{dt} + \frac{T}{\mu}\frac{d\mu}{dt} + \frac{2 \mu m_p}{3 \rho k_B}\left({\mathcal H} - \Lambda\right).
\label{eq:tev}
\end{align}
The first term describes the adiabatic cooling from the overall expansion of the universe, while the second term accounts for adiabatic cooling/heating from structure
formation as elements expand/contract, and the third term results from changes in the mean mass per particle. The final term takes into account other
relevant heating/cooling processes; in our case, the most important processes included in this term are photoionization
heating of HI, and Compton cooling. 

In general, one needs to solve for the thermal evolution of the IGM gas in conjunction with equations describing the evolving
ionization state and density of each gas element. Here, we can gain some insight by assuming the gas is highly ionized and in ionization
equilibrium {\em after it is heated to a temperature $T_{\rm reion}$ during reionization}, and by including only the dominant heating/cooling processes: adiabatic heating/cooling, Compton cooling, and
HI photoheating. In this case, we can approximate Eq. \ref{eq:tev} by \cite{Lidz:2014jxa}:
\begin{align}
\frac{dT}{dt} =& -2 H T + \frac{2 T}{3 (1+\delta)}\frac{d\delta}{dt} + \frac{\alpha_0 \bar{n}_e E_J}{3 (1 + \chi_{\rm He}) k_B} \left(\frac{T}{10^4 K}\right)^{-0.7} (1 + \delta) \nonumber \\
 & +  \frac{4}{3}\frac{\sigma_T a_{\rm rad} T_\gamma^4}{m_e c} \left(T_\gamma - T \right).
\label{eq:tev_approx}
\end{align}
We adopt here the case-A recombination rate and approximate it as $\alpha_A = 4.2 \times 10^{-13} (T/10^4 K)^{-0.7}$ cm$^3 s^{-1}$;
$\alpha_0$ denotes the recombination coefficient of hydrogen at $T=10^4$ K. Here $\bar{n}_e$ is the cosmic mean electron density, $\chi_{\rm He} = n_{\rm He}/n_{\rm H}$ is
the helium fraction, $E_J$ is the average energy input per photoionization after reionization, $\sigma_T$ is the Thomson scattering cross section,
$T_\gamma$ is the CMB temperature, and $a_{\rm rad} T_\gamma^4$ is the energy density in the CMB. 

Furthermore, we can assume a solution of
the form $T=T_0 (1+\delta)^{\gamma-1}$ and linearize ($T \approx T_0 [1 + (\gamma-1) \delta]$) to find approximate evolution equations
for each of $T_0$ and $\gamma-1$ (\cite{Hui:1997dp,Lidz:2014jxa}).  This linear approach is fruitful because hydrodynamic simulations show that the temperature-density
relation -- for gas elements that reionize at a given redshift -- is well approximated by a power-law.\footnote{The power law relation is only a good description for the low density IGM where shock-heating
is relatively unimportant, as are many cooling processes (which are efficient only at higher densities). The Ly-$\alpha$ forest at the redshifts of interest is insensitive to this rather overdense gas since it 
produces saturated absorption.} Thus, even though linear theory provides an imperfect description over the full range in density probed by the Ly-$\alpha$ forest, it suffices
to determine the slope and amplitude of this power-law \cite{Hui:1997dp}.
Here we follow the Appendix of Lidz \& Malloy (2014) \cite{Lidz:2014jxa}. Linearizing then gives:
\beqa
\frac{d(a^2 T_0)}{da} =  {\mathcal A} a^{0.9} (a^2 T_0)^{-0.7} - {\mathcal B} a^{-7/2} (a^2 T_0) + {\mathcal B} T_\gamma(0) a^{-5/2
},
\label{eq:tzero_eq}
\eeqa
and
\begin{align}
\frac{d(\gamma-1)}{da} =& \left[\frac{2}{3} - (\gamma-1)\right] \frac{1}{a} + {\mathcal A} a^{0.9} (a^2 T_0)^{-1.7} \left[1 - 1.7 (
\gamma-1) \right] \nonumber \\
& - \frac{{\mathcal B} T_\gamma(0)}{a^2 T_0} a^{-5/2} (\gamma-1).
\label{eq:gamma_eos}
\end{align}
These equations are valid to linear order in $\delta$.  In the above equations, we have used two constants ${\mathcal A}$ and ${\mathcal B}$ defined by:
\beqa
{\mathcal A} = (10^4 K)^{0.7} \frac{\alpha_0 \bar{n}_e(0) E_J}{3 (1 + \chi_{\rm He}) k_B H_0 \sqrt{\Omega_m}},
\label{eq:adef}
\eeqa
and
\beqa
{\mathcal B} = \frac{1}{H_0 \sqrt{\Omega_m} t_{\rm Comp}(0)}; \quad
t_{\rm Comp} = \frac{3 m_e c}{4 \sigma_T a_{\rm rad} T_\gamma^4}.
\label{eq:bdef}
\eeqa

\begin{figure}[b]
\bc
\includegraphics[scale=0.65]{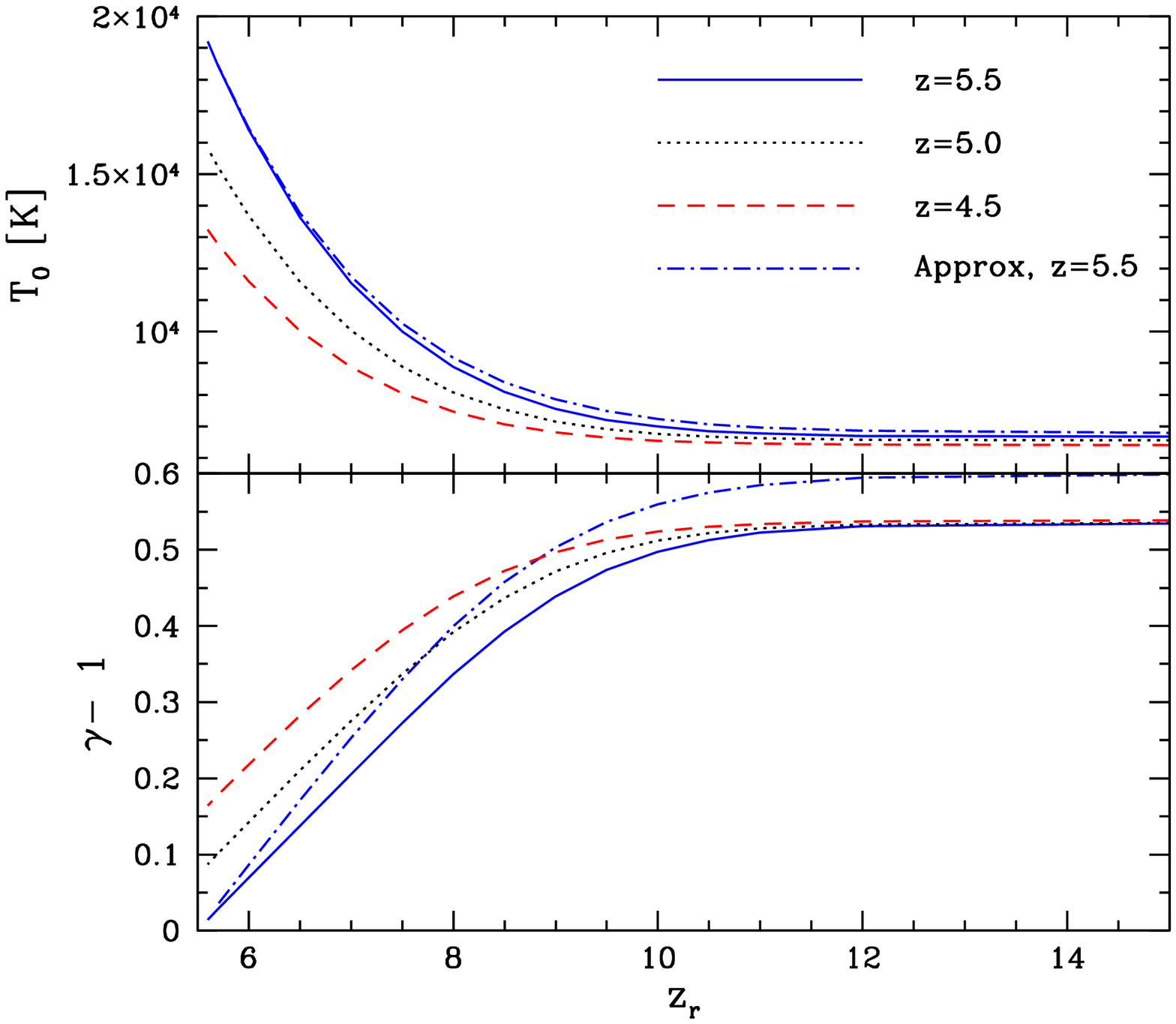}
\caption{Temperature of the IGM for gas elements at each of $z=4.5, 5.0$, and $z=5.5$ as a function of the redshift at which the gas is reionized. The top panel shows
the temperature at mean density ($T_0$), while the bottom panel shows the slope of the temperature-density relation, $\gamma$.   In each panel, the blue dot-dashed line
shows the  approximate evolution predicted by Eqs. \ref{eq:tzero_eq} and \ref{eq:gamma_eos}, while the other lines include further relevant cooling processes and adopt
the Zel'dovich approximation to follow the density evolution, rather than assuming linear theory. In each case, we assume that the gas is heated to a fixed temperature
at reionization, $T_r = 2 \times 10^4$ K, and assume a (hardened) spectral index of $\alpha=1.5$ for computing the post-reionization photoheating. Note that although we assume
that all gas that reionizes at a given redshift lands on a well-defined temperature-density relation, this will {\em not typically} be a good description once we account for the spread
in reionization redshifts across the universe. Adapted from \cite{Lidz:2014jxa}.}
\label{fig:therm_hist}
\ec
\end{figure}

The solution of Eq. \ref{eq:tzero_eq} has a closed-form analytic solution given in \cite{Lidz:2014jxa}, but Eq. \ref{eq:gamma_eos} needs to be solved numerically.
Other convenient approximations -- beyond the linear perturbation treatment of Eqs. \ref{eq:tzero_eq} and \ref{eq:gamma_eos}  --  are to solve for the density evolution assuming either spherical collapse \cite{Furlanetto:2009kr} or the Zel'dovich approximation \cite{Hui:1997dp,Lidz:2014jxa}. Fig. \ref{fig:therm_hist} compares the linear perturbation approach (with the above heating/cooling processes) to results that track the density evolution (of many tracer gas elements)
using the Zel'dovich approximation \cite{1970A&A.....5...84Z}, while also adopting a more complete treatment of heating/cooling processes. Specifically, the figure shows how the temperature of gas elements at each of
$z=4.5, 5.0$, and $z=5.5$ vary with the redshift at which they are reionized, $z_r$. In general, the gas elements that reionize at very high redshift are at smaller temperatures than gas that reionized
recently: this is because gas that reionizes early has longer to cool and reaches a lower temperature than recently ionized gas. However,  the temperature becomes {\em insensitive} to the reionization redshift for gas that is reionized at sufficiently high redshift: in this model, the temperature at mean density for all gas with $z_{\rm reion} \geq10$ is $T_0 = 6,700$ K at $z=5.5$, irrespective of the reionization redshift. This occurs
because Compton cooling is efficient enough at $z \geq 10$ to erase any memory of prior photo-heating. In any case, one can see that the post-reionization IGM temperature generally retains some memory of
prior photo-heating during reionization: if the temperature can be accurately measured, this can be turned into a probe of the EoR itself.\footnote{Note that although the memory of the photoheating during
hydrogen reionization further fades towards lower redshift, photoheating from HeII reionization should start to impact the gas  significantly (e.g. \cite{McQuinn:2008am}.)} The dot-dashed line in Fig. \ref{fig:therm_hist} also illustrates
that the simplified expression in Fig. \ref{eq:tzero_eq} provides a good match to the more detailed models (other curves), although it predicts a more rapid steepening in $\gamma$ with
redshift -- and reaches a higher asymptotic value of $\gamma$ -- than found in the Zel'dovich approximation calculations.  
Note that quantitatively all the trends shown here are somewhat sensitive to the temperature reached at reionization (here $T_r = 2 \times 10^4$ K) and the spectral index of the ionizing radiation after reionization (here $\alpha=1.5$),
which sets the amount of residual post-reionization photo-heating (see e.g. \cite{Lidz:2014jxa} for more details).

\begin{figure}[b]
\sidecaption
\includegraphics[scale=0.65]{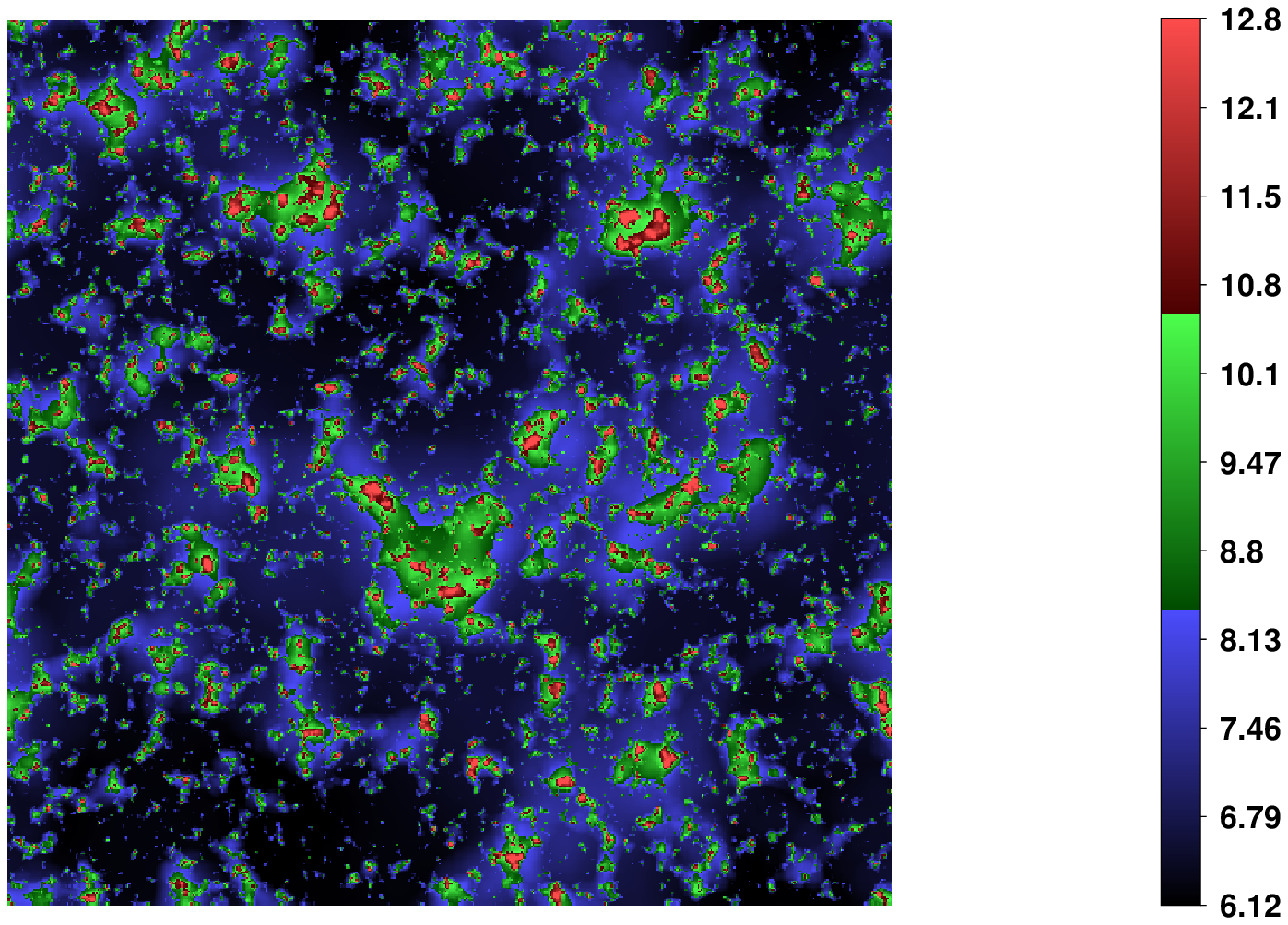}
\includegraphics[scale=0.65]{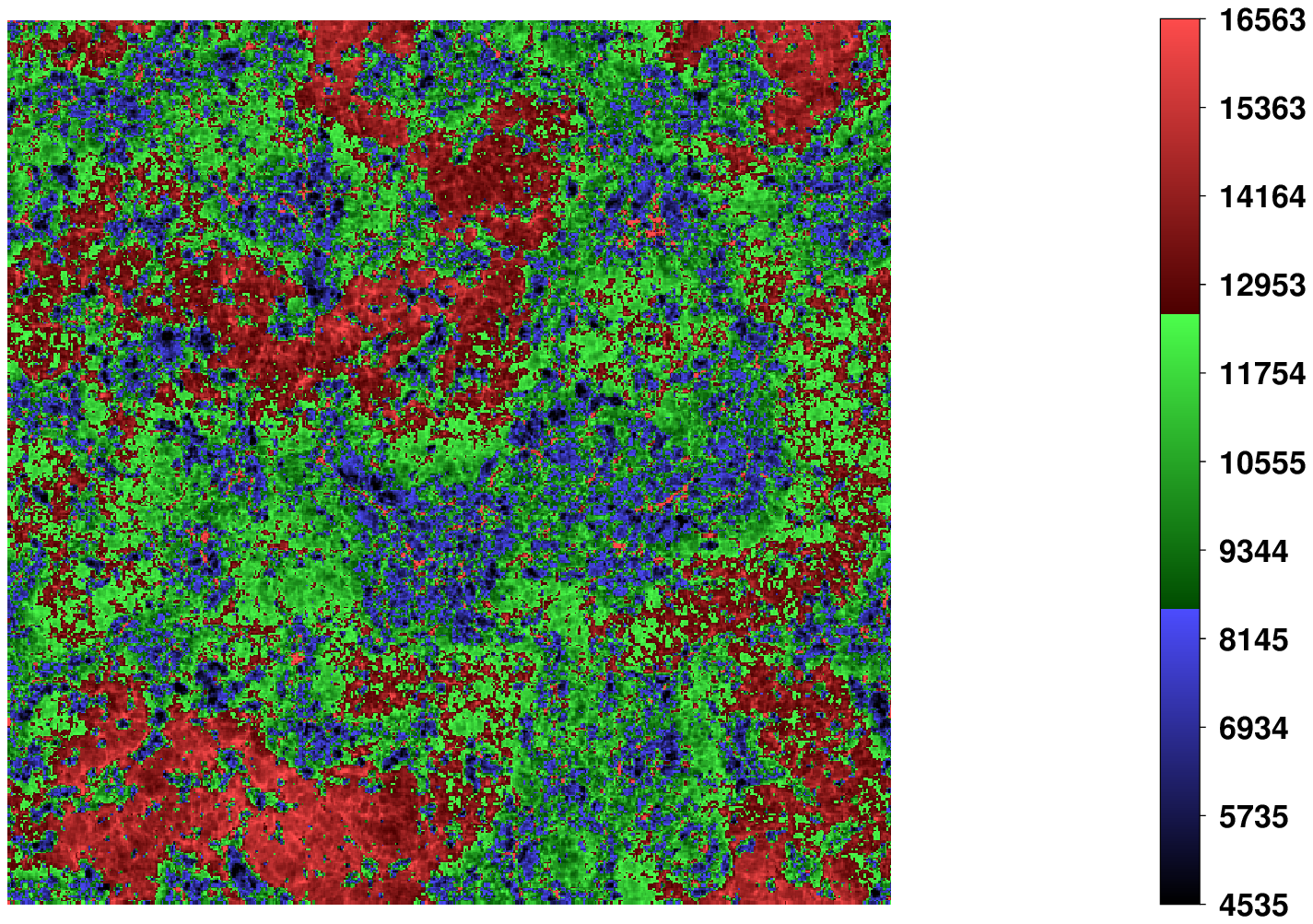}
\caption{Inhomogeneous thermal state of the IGM. The two panels illustrate how spatial variations in the redshift
of reionization may lead to temperature inhomogeneities after reionization. The panels are from narrow slices ($0.25$ Mpc/$h$ thick)
through a semi-analytic reionization model in which reionization completes at $z=5.8$. Each slice is $130$ Mpc/$h$ on
a side. {\em Top panel:} The reionization redshifts across the simulation; the red regions have the highest reionization redshifts
across the slice, while the dark regions reionize at lower redshift. {\em Bottom panel:}
The $z=5.5$ temperature through the same slice as in the top panel. The red areas in this panel show the
hottest locations in the slice, and correspond to the dark regions in the top panel that
are reionized late. The dark blue regions in the temperature slice, on the other hand, are the coolest regions that reionized
first. From \cite{Lidz:2014jxa}.}
\label{fig:tslice_lowz}
\end{figure}

In the model of Figure \ref{fig:therm_hist}, the temperature of each gas element in the IGM is entirely specified by its reionization redshift. It is then interesting to consider the spatial
variations in the temperature of the IGM that might result from fluctuations in the timing of reionization across the universe. Fig. \ref{fig:tslice_lowz} shows a slice through a simulated
model of the $z \sim 5.5$ IGM that incorporates spatial variations in the redshift of reionization. In this particular model, reionization completes at redshift $z=5.8$ -- just
slightly higher than the redshift considered ($z=5.5$) -- but some regions are reionized significantly earlier. In particular, the top panel of the figure shows the reionization redshifts
for different regions of the universe extracted from a narrow slice through the simulation volume, while the bottom panel shows the temperature across the same slice.
One can see that the temperature field has interesting large-scale fluctuations, with regions that ionize late having high temperatures compared to locations that ionize early on (see also \cite{Trac:2008yz,Furlanetto:2009kr}).
In models in which reionization occurs earlier (on average), the average temperature should be smaller as should the spatial fluctuations in the temperature, since efficient
Compton cooling at $z \geq 10$ makes the post-reionization gas insensitive to the reionization redshift when it becomes reionized this early. If the temperature of the IGM
can be measured accurately enough from Ly-$\alpha$ forest data at high redshift \cite{Becker11,Lidz:2014jxa}, this can be used to constrain the reionization history of the universe and possibly to
extract information about the spatial fluctuations around the average reionization history.

\section{The Gas Density Distribution}
\label{sec:gas_pdf}

The probability distribution of the gas density field is an important quantity for making contact with observations of the $z \sim 6$ Ly-$\alpha$ forest and other probes of the high redshift
IGM. In  particular, the large optical depth to Ly-$\alpha$ absorption at $z \sim 6$ implies that the transmission through the high-z
Ly-$\alpha$ forest is sensitive to the rare, low-density tail of the gas density probability distribution function (see e.g. \cite{Oh:2004rm}). In addition, models of the gas density distribution
are often used to estimate clumping factors and the mean-free path between dense clumps, and so this quantity is broadly important for understanding the IGM
during (and after) reionization.

The gas density distribution is itself sensitive to the reionization history of the universe. Photoheating during reionization leads to enhanced pressure gradients in the 
gas, and heated regions expand in response, on roughly the sound crossing timescale. In recently heated regions of the universe, however, the gas will not have had time to fully respond to the prior heating. This can partly be quantified using the ``filtering scale'' of \cite{Gnedin:1997td}; while instructive, the linear analysis considered in that work will not, however, fully capture the impact of pressure smoothing.
Note also that before reionization completes, there will still be some mostly cold and neutral patches in the IGM that have experienced only small amounts of ``pre-heating'' from, e.g., exposure to low-levels of X-ray heating (e.g. \cite{Oh:2003pm}). In the cold or recently heated regions of the IGM, the gas should trace the dark matter more closely than in high temperature regions that experienced earlier heating.
This is the same ``positive feedback'' effect of photoheating we discussed in \S \ref{sec:clumping} in the context of clumping factors; here we focus on the impact of photoheating on the full
volume-averaged density PDF \cite{Pawlik:2008mr}.  Along with preventing gas from collapsing into small halos, and expelling some gas from collapsed halos, Jeans smoothing should prevent
the low density tail of the PDF from extending as prominently to low density as it would otherwise. This might impact the transmission in the $z \sim 6$ Ly-$\alpha$ forest, which is sensitive to the low density tail of the PDF. However, Fig. 3 of \cite{Pawlik:2008mr} suggests that the low density tail depends weakly on the reheating history.

A commonly used fitting function to describe the volume averaged PDF of the density field, introduced in \cite{MiraldaEscude:1998qs}, is :
\beqa
P_V(\Delta) d\Delta = A\ {\rm exp} \bigg[-\frac{\left(\Delta^{-2/3} - C_0\right)^2}{8 \delta_0^2/9}\bigg] \Delta^{-\beta} d \Delta.
\label{eq:pdf_delta}
\eeqa
Here $\Delta=\rho_g/\avg{\rho_g}$ is the gas density in units of the cosmic mean gas density. More recent work has updated the original results from \cite{MiraldaEscude:1998qs} by determining the
best fit parameters $(A,C_0, \beta)$ using improved numerical simulations \cite{Pawlik:2008mr}. These authors find $A=3.038$, $\delta_0 = 1.477$, $\beta = 3.380$, $C_0=-0.932$ at $z=6$ after fitting their
simulation results for densities over the range of $0.1 \leq \Delta \leq 100$. These results assume a heating history in which the universe is instantaneously reionized at $z_r=9$, and the gas
is heated at reionization to a temperature of $T_r \sim 10^4$ K. Note that this fitting formula provides a good, but imperfect fit to the simulation results (see Fig. A1 of \cite{Pawlik:2008mr} and also \cite{2009MNRAS.398L..26B}, who obtain slightly different results for the PDF). As mentioned in the clumping factor section, these results assume the optically thin limit and ignore the impact of self-shielding on the photo-evaporation process; further
radiation-hydrodynamical modeling along the lines of \cite{Shapiro:2003gxa} might be valuable here. Furthermore, the simulations neglect the impact of the X-ray pre-heating effects we mentioned previously.
In any case, the gas density distribution model here is a useful tool for analytic and semi-analytic estimates.

\section{Conclusion}
\label{sec:conclusion}

The observational prospects for improving our understanding of the EoR over the next few years are outstanding,  and a great deal of theoretical work is required in the meantime to best prepare for the upcoming measurements. A broad range of different probes, across a wide variety of wavebands, should soon provide increasingly detailed information regarding our cosmic dawn era, including
observations of: the redshifted 21 cm line, large-scale anisotropies in the polarization of the CMB, small-scale CMB fluctuations, LBG surveys, LAE emitter surveys, GRB optical afterglows, quasar absorption
spectra, fluctuations in the infrared background, improved measurements of local group dwarf galaxies, and other probes. Although these observations are all promising, the signatures of reionization
are often subtle and a comparison with detailed theoretical models will be required to interpret these measurements. 

Modeling the IGM during reionization is challenging, primarily because the interplay between the first luminous sources and the surrounding intergalactic gas involves a huge dynamic range in scale; in principle, spatial scales spanning the entire range from the size of individual stars or accreting black holes, to the hundreds of co-moving megaparsec scales required to obtain a representative sample of the large ionized regions during reionization, are relevant. In order to face this challenge, flexible models with sub-grid prescriptions for the sources and sinks of ionizing photons are necessary. These models must be calibrated against existing observations, and using smaller scale simulations that provide a more detailed treatment of the underlying physical processes. In conjunction with these studies, analytic and semi-analytic models help to identify and isolate the essential physics involved, and allow a fast exploration of parameter space. In this chapter, we summarized a few
aspects of reionization-era phenomenology, often emphasizing the importance of spatial variations in the properties of the IGM during reionization. 

We are optimistic that the reionization models described here will soon be improved and refined, in part by comparing with existing and near future observations, and that we will then be in good
shape to interpret future, more detailed, data sets.
These studies should allow us to determine the redshift evolution of the volume-averaged ionization fraction, and the size distribution of the ionized regions at different stages of the reionization process. This will in turn help determine the properties of the first luminous sources in our universe and the nature of early phases of structure formation, and also help in understanding the impact of these sources on subsequent galaxy formation.

\begin{acknowledgement}

AL acknowledges support from NASA grant NNX12AC97G and from the NSF through grant AST-1109156.  I thank the editor of this volume, Andrei Mesinger, for very helpful feedback on a draft version of
this chapter. I also thank my collaborators on reionization-related projects over the years for joint work on some of the topics discussed in this chapter, especially: Mark Dijkstra, Suvendra Dutta, 
Claude-Andr\'e Faucher-Gigu\`ere, Steve Furlanetto, Lars Hernquist, Lam Hui, Matt Malloy, Matt McQuinn, Peng Oh, Jessie Taylor, Oliver Zahn, Fangzhou Zhu, and Matias Zaldarriaga.

\end{acknowledgement}

\bibliographystyle{unsrt}
\bibliography{references}

\begin{thebibliography}{100}

\bibitem{Bond:1995yt}
J.~Richard Bond, Lev Kofman, and Dmitri Pogosyan.
\newblock {How filaments are woven into the cosmic web}.
\newblock {\em Nature}, 380:603--606, 1996.

\bibitem{McDonald:2004eu}
Patrick McDonald et~al.
\newblock {The Lyman-alpha forest power spectrum from the Sloan Digital Sky
  Survey}.
\newblock {\em Astrophys.J.Suppl.}, 163:80--109, 2006.

\bibitem{Zaldarriaga:2008ap}
Matias Zaldarriaga, Loris Colombo, Eiichiro Komatsu, Adam Lidz, Michael
  Mortonson, et~al.
\newblock {CMBPol Mission Concept Study: Reionization Science with the Cosmic
  Microwave Background}.
\newblock 2008.

\bibitem{Zahn:2005fn}
Oliver Zahn, Matias Zaldarriaga, Lars Hernquist, and Matthew McQuinn.
\newblock {The Influence of non-uniform reionization on the CMB}.
\newblock {\em Astrophys.J.}, 630:657--666, 2005.

\bibitem{McQuinn:2005ce}
Matthew McQuinn, Steven~R. Furlanetto, Lars Hernquist, Oliver Zahn, and Matias
  Zaldarriaga.
\newblock {The Kinetic Sunyaev-Zel'dovich effect from reionization}.
\newblock {\em Astrophys.J.}, 630:643--656, 2005.

\bibitem{2013fgu..book.....L}
A.~{Loeb} and S.~R. {Furlanetto}.
\newblock {\em {The First Galaxies in the Universe}}.
\newblock 2013.

\bibitem{Fan:2005es}
Xiao-Hui Fan, Michael~A. Strauss, Robert~H. Becker, Richard~L. White, James~E.
  Gunn, et~al.
\newblock {Constraining the evolution of the ionizing background and the epoch
  of reionization with z~6 quasars. 2. a sample of 19 quasars}.
\newblock {\em Astron.J.}, 132:117--136, 2006.

\bibitem{Mortlock11}
D.~J. {Mortlock}, S.~J. {Warren}, B.~P. {Venemans}, M.~{Patel}, P.~C. {Hewett},
  R.~G. {McMahon}, C.~{Simpson}, T.~{Theuns}, E.~A. {Gonz{\'a}les-Solares},
  A.~{Adamson}, S.~{Dye}, N.~C. {Hambly}, P.~{Hirst}, M.~J. {Irwin},
  E.~{Kuiper}, A.~{Lawrence}, and H.~J.~A. {R{\"o}ttgering}.
\newblock {A luminous quasar at a redshift of z = 7.085}.
\newblock {\em \nat}, 474:616--619, June 2011.

\bibitem{Konno:2014fja}
Akira Konno, Masami Ouchi, Yoshiaki Ono, Kazuhiro Shimasaku, Takatoshi Shibuya,
  et~al.
\newblock {Accelerated Evolution of Ly$\alpha$ Luminosity Function at
  $\textit{z} \gtrsim 7$ Revealed by the Subaru Ultra-Deep Survey for
  Ly$\alpha$ Emitters at $\textit{z}=7.3$}.
\newblock 2014.

\bibitem{Bouwens:2014fua}
R.J. Bouwens, G.D. Illingworth, P.A. Oesch, M.~Trenti, I.~Labbe', et~al.
\newblock {UV Luminosity Functions at redshifts z~4 to z~10: 10000 Galaxies
  from HST Legacy Fields}.
\newblock 2014.

\bibitem{Totani:2013nra}
Tomonori Totani, Kentaro Aoki, Takashi Hattori, George Kosugi, Yuu Niino,
  et~al.
\newblock {Probing Intergalactic Neutral Hydrogen by the Lyman Alpha Red
  Damping Wing of Gamma-Ray Burst 130606A Afterglow Spectrum at z = 5.913}.
\newblock 2013.

\bibitem{Komatsu11}
E.~{Komatsu}, K.~M. {Smith}, J.~{Dunkley}, C.~L. {Bennett}, B.~{Gold},
  G.~{Hinshaw}, N.~{Jarosik}, D.~{Larson}, M.~R. {Nolta}, L.~{Page}, D.~N.
  {Spergel}, M.~{Halpern}, R.~S. {Hill}, A.~{Kogut}, M.~{Limon}, S.~S. {Meyer},
  N.~{Odegard}, G.~S. {Tucker}, J.~L. {Weiland}, E.~{Wollack}, and E.~L.
  {Wright}.
\newblock {Seven-year Wilkinson Microwave Anisotropy Probe (WMAP) Observations:
  Cosmological Interpretation}.
\newblock {\em \apjs}, 192:18, February 2011.

\bibitem{Zahn12}
O.~{Zahn}, C.~L. {Reichardt}, L.~{Shaw}, A.~{Lidz}, K.~A. {Aird}, B.~A.
  {Benson}, L.~E. {Bleem}, J.~E. {Carlstrom}, C.~L. {Chang}, H.~M. {Cho}, T.~M.
  {Crawford}, A.~T. {Crites}, T.~{de Haan}, M.~A. {Dobbs}, O.~{Dor{\'e}},
  J.~{Dudley}, E.~M. {George}, N.~W. {Halverson}, G.~P. {Holder}, W.~L.
  {Holzapfel}, S.~{Hoover}, Z.~{Hou}, J.~D. {Hrubes}, M.~{Joy}, R.~{Keisler},
  L.~{Knox}, A.~T. {Lee}, E.~M. {Leitch}, M.~{Lueker}, D.~{Luong-Van}, J.~J.
  {McMahon}, J.~{Mehl}, S.~S. {Meyer}, M.~{Millea}, J.~J. {Mohr}, T.~E.
  {Montroy}, T.~{Natoli}, S.~{Padin}, T.~{Plagge}, C.~{Pryke}, J.~E. {Ruhl},
  K.~K. {Schaffer}, E.~{Shirokoff}, H.~G. {Spieler}, Z.~{Staniszewski}, A.~A.
  {Stark}, K.~{Story}, A.~{van Engelen}, K.~{Vanderlinde}, J.~D. {Vieira}, and
  R.~{Williamson}.
\newblock {Cosmic Microwave Background Constraints on the Duration and Timing
  of Reionization from the South Pole Telescope}.
\newblock {\em \apj}, 756:65, September 2012.

\bibitem{Parsons:2013dwa}
Aaron~R. Parsons, Adrian Liu, James~E. Aguirre, Zaki~S. Ali, Richard~F.
  Bradley, et~al.
\newblock {New Limits on 21cm EoR From PAPER-32 Consistent with an X-Ray Heated
  IGM at z=7.7}.
\newblock {\em Astrophys.J.}, 788:106, 2014.

\bibitem{Paciga:2013fj}
Gregory Paciga, Joshua Albert, Kevin Bandura, Tzu-Ching Chang, Yashwant Gupta,
  et~al.
\newblock {A refined foreground-corrected limit on the HI power spectrum at
  z=8.6 from the GMRT Epoch of Reionization Experiment}.
\newblock 2013.

\bibitem{Dillon:2013rfa}
Joshua~S. Dillon, Adrian Liu, Christopher~L. Williams, Jacqueline~N. Hewitt,
  Max Tegmark, et~al.
\newblock {Overcoming real-world obstacles in 21 cm power spectrum estimation:
  A method demonstration and results from early Murchison Widefield Array
  data}.
\newblock {\em Phys.Rev.}, D89:023002, 2014.

\bibitem{Yatawatta13}
S.~{Yatawatta}, A.~G. {de Bruyn}, M.~A. {Brentjens}, P.~{Labropoulos}, V.~N.
  {Pandey}, S.~{Kazemi}, S.~{Zaroubi}, L.~V.~E. {Koopmans}, A.~R. {Offringa},
  V.~{Jeli{\'c}}, O.~{Martinez Rubi}, V.~{Veligatla}, S.~J. {Wijnholds}, W.~N.
  {Brouw}, G.~{Bernardi}, B.~{Ciardi}, S.~{Daiboo}, G.~{Harker}, G.~{Mellema},
  J.~{Schaye}, R.~{Thomas}, H.~{Vedantham}, E.~{Chapman}, F.~B. {Abdalla},
  A.~{Alexov}, J.~{Anderson}, I.~M. {Avruch}, F.~{Batejat}, M.~E. {Bell}, M.~R.
  {Bell}, M.~{Bentum}, P.~{Best}, A.~{Bonafede}, J.~{Bregman}, F.~{Breitling},
  R.~H. {van de Brink}, J.~W. {Broderick}, M.~{Br{\"u}ggen}, J.~{Conway},
  F.~{de Gasperin}, E.~{de Geus}, S.~{Duscha}, H.~{Falcke}, R.~A. {Fallows},
  C.~{Ferrari}, W.~{Frieswijk}, M.~A. {Garrett}, J.~M. {Griessmeier}, A.~W.
  {Gunst}, T.~E. {Hassall}, J.~W.~T. {Hessels}, M.~{Hoeft}, M.~{Iacobelli},
  E.~{Juette}, A.~{Karastergiou}, V.~I. {Kondratiev}, M.~{Kramer},
  M.~{Kuniyoshi}, G.~{Kuper}, J.~{van Leeuwen}, P.~{Maat}, G.~{Mann}, J.~P.
  {McKean}, M.~{Mevius}, J.~D. {Mol}, H.~{Munk}, R.~{Nijboer}, J.~E. {Noordam},
  M.~J. {Norden}, E.~{Orru}, H.~{Paas}, M.~{Pandey-Pommier}, R.~{Pizzo}, A.~G.
  {Polatidis}, W.~{Reich}, H.~J.~A. {R{\"o}ttgering}, J.~{Sluman},
  O.~{Smirnov}, B.~{Stappers}, M.~{Steinmetz}, M.~{Tagger}, Y.~{Tang},
  C.~{Tasse}, S.~{ter Veen}, R.~{Vermeulen}, R.~J. {van Weeren}, M.~{Wise},
  O.~{Wucknitz}, and P.~{Zarka}.
\newblock {Initial deep LOFAR observations of epoch of reionization windows. I.
  The north celestial pole}.
\newblock {\em \aap}, 550:A136, February 2013.

\bibitem{Palanque-Delabrouille:2014jca}
Nathalie Palanque-Delabrouille, Christophe Yche, Julien Lesgourgues, Graziano
  Rossi, Arnaud Borde, et~al.
\newblock {Constraint on neutrino masses from SDSS-III/BOSS Ly$\alpha$ forest
  and other cosmological probes}.
\newblock 2014.

\bibitem{Cen:1994da}
Ren-yue Cen, Jordi Miralda-Escude, Jeremiah~P. Ostriker, and Michael Rauch.
\newblock {Gravitational collapse of small scale structure as the origin of the
  Lyman alpha forest}.
\newblock {\em Astrophys.J.}, 437:L9, 1994.

\bibitem{Hernquist:1995uma}
Lars Hernquist, Neal Katz, David~H. Weinberg, and Jordi Miralda-Escude.
\newblock {The Lyman alpha forest in the cold dark matter model}.
\newblock {\em Astrophys.J.}, 457:L51, 1996.

\bibitem{MiraldaEscude:1995bu}
Jordi Miralda-Escude, Ren-yue Cen, Jeremiah~P. Ostriker, and Michael Rauch.
\newblock {The Lyman alpha forest from gravitational collapse in the CDM +
  Lambda Model}.
\newblock {\em Astrophys.J.}, 471:582, 1996.

\bibitem{Hui:1996fh}
Lam Hui, Nickolay~Y. Gnedin, and Yu~Zhang.
\newblock {The Statistics of density peaks and the column density distribution
  of the Lyman-alpha forest}.
\newblock {\em Astrophys.J.}, 486:599, 1997.

\bibitem{Viel:2004bf}
Matteo Viel, Martin~G. Haehnelt, and Volker Springel.
\newblock {Inferring the dark matter power spectrum from the Lyman-alpha forest
  in high-resolution QSO absorption spectra}.
\newblock {\em Mon.Not.Roy.Astron.Soc.}, 354:684, 2004.

\bibitem{McQuinn:2008am}
Matthew McQuinn, Adam Lidz, Matias Zaldarriaga, Lars Hernquist, Philip~F.
  Hopkins, et~al.
\newblock {HeII Reionization and its Effect on the IGM}.
\newblock {\em Astrophys.J.}, 694:842--866, 2009.

\bibitem{Compostella:2013zya}
Michele Compostella, Sebastiano Cantalupo, and Cristiano Porciani.
\newblock {The imprint of inhomogeneous HeII reionization on the HI and HeII
  Ly-alpha forest}.
\newblock {\em Mon.Not.Roy.Astron.Soc.}, 435:3169--3190, 2013.

\bibitem{1987ApJ...321L.107S}
P.~R. {Shapiro} and M.~L. {Giroux}.
\newblock {Cosmological H II regions and the photoionization of the
  intergalactic medium}.
\newblock {\em \apjl}, 321:L107--L112, October 1987.

\bibitem{1999ApJ...514..648M}
P.~{Madau}, F.~{Haardt}, and M.~J. {Rees}.
\newblock {Radiative Transfer in a Clumpy Universe. III. The Nature of
  Cosmological Ionizing Sources}.
\newblock {\em \apj}, 514:648--659, April 1999.

\bibitem{2008ApJ...688...85F}
C.-A. {Faucher-Gigu{\`e}re}, A.~{Lidz}, L.~{Hernquist}, and M.~{Zaldarriaga}.
\newblock {Evolution of the Intergalactic Opacity: Implications for the
  Ionizing Background, Cosmic Star Formation, and Quasar Activity}.
\newblock {\em \apj}, 688:85--107, November 2008.

\bibitem{1994MNRAS.266..343M}
J.~{Miralda-Escud{\'e}} and M.~J. {Rees}.
\newblock {Reionization and thermal evolution of a photoionized intergalactic
  medium.}
\newblock {\em \mnras}, 266:343--352, January 1994.

\bibitem{Hui:1997dp}
Lam Hui and Nickolay~Y. Gnedin.
\newblock {Equation of state of the photoionized intergalactic medium}.
\newblock {\em Mon.Not.Roy.Astron.Soc.}, 292:27, 1997.

\bibitem{Gnedin:1997td}
Nickolay~Y. Gnedin and Lam Hui.
\newblock {Probing the universe with the Lyman alpha forest: 1. Hydrodynamics
  of the low density IGM}.
\newblock {\em Mon.Not.Roy.Astron.Soc.}, 296:44--55, 1998.

\bibitem{2009ApJ...705L.113P}
J.~X. {Prochaska}, G.~{Worseck}, and J.~M. {O'Meara}.
\newblock {A Direct Measurement of the Intergalactic Medium Opacity to H I
  Ionizing Photons}.
\newblock {\em \apjl}, 705:L113--L117, November 2009.

\bibitem{2011ASL.....4..228T}
H.~Y. {Trac} and N.~Y. {Gnedin}.
\newblock {Computer Simulations of Cosmic Reionization}.
\newblock {\em Advanced Science Letters}, 4:228--243, February 2011.

\bibitem{2009MNRAS.400.1283I}
I.~T. {Iliev}, D.~{Whalen}, G.~{Mellema}, K.~{Ahn}, S.~{Baek}, N.~Y. {Gnedin},
  A.~V. {Kravtsov}, M.~{Norman}, M.~{Raicevic}, D.~R. {Reynolds}, D.~{Sato},
  P.~R. {Shapiro}, B.~{Semelin}, J.~{Smidt}, H.~{Susa}, T.~{Theuns}, and
  M.~{Umemura}.
\newblock {Cosmological radiative transfer comparison project - II. The
  radiation-hydrodynamic tests}.
\newblock {\em \mnras}, 400:1283--1316, December 2009.

\bibitem{Zahn:2006sg}
Oliver Zahn, Adam Lidz, Matthew McQuinn, Suvendra Dutta, Lars Hernquist, et~al.
\newblock {Simulations and Analytic Calculations of Bubble Growth During
  Hydrogen Reionization}.
\newblock {\em Astrophys.J.}, 654:12--26, 2006.

\bibitem{2011MNRAS.411..955M}
A.~{Mesinger}, S.~{Furlanetto}, and R.~{Cen}.
\newblock {21CMFAST: a fast, seminumerical simulation of the high-redshift
  21-cm signal}.
\newblock {\em \mnras}, 411:955--972, February 2011.

\bibitem{Furlanetto:2004nh}
Steven Furlanetto, Matias Zaldarriaga, and Lars Hernquist.
\newblock {The Growth of HII regions during reionization}.
\newblock {\em Astrophys.J.}, 613:1--15, 2004.

\bibitem{1991ApJ...379..440B}
J.~R. {Bond}, S.~{Cole}, G.~{Efstathiou}, and N.~{Kaiser}.
\newblock {Excursion set mass functions for hierarchical Gaussian
  fluctuations}.
\newblock {\em \apj}, 379:440--460, October 1991.

\bibitem{McQuinn:2012bq}
Matthew McQuinn.
\newblock {Constraints on X-ray Emissions from the Reionization Era}.
\newblock {\em Mon.Not.Roy.Astron.Soc.}, 426:1349--1360, 2012.

\bibitem{Haiman:1996ht}
Zoltan Haiman and Abraham Loeb.
\newblock {Signatures of stellar reionization of the universe}.
\newblock {\em Astrophys.J.}, 483:21, 1997.

\bibitem{Birnboim:2003xa}
Yuval Birnboim and Avishai Dekel.
\newblock {Virial shocks in galactic haloes?}
\newblock {\em Mon.Not.Roy.Astron.Soc.}, 345:349--364, 2003.

\bibitem{Keres:2004cq}
Dusan Keres, Neal Katz, David~H. Weinberg, and Romeel Dave.
\newblock {How do galaxies get their gas?}
\newblock {\em Mon.Not.Roy.Astron.Soc.}, 363:2--28, 2005.

\bibitem{Barkana:2000fd}
Rennan Barkana and Abraham Loeb.
\newblock {In the beginning: The First sources of light and the reionization of
  the Universe}.
\newblock {\em Phys.Rept.}, 349:125--238, 2001.

\bibitem{Dekel86}
A.~{Dekel} and J.~{Silk}.
\newblock {The origin of dwarf galaxies, cold dark matter, and biased galaxy
  formation}.
\newblock {\em \apj}, 303:39--55, April 1986.

\bibitem{Efstathiou92}
G.~{Efstathiou}.
\newblock {Suppressing the formation of dwarf galaxies via photoionization}.
\newblock {\em \mnras}, 256:43P--47P, May 1992.

\bibitem{Thoul96}
A.~A. {Thoul} and D.~H. {Weinberg}.
\newblock {Hydrodynamic Simulations of Galaxy Formation. II. Photoionization
  and the Formation of Low-Mass Galaxies}.
\newblock {\em \apj}, 465:608, July 1996.

\bibitem{Barkana:1999apa}
Rennan Barkana and Abraham Loeb.
\newblock {The photoevaporation of dwarf galaxies during reionization}.
\newblock {\em Astrophys.J.}, 523:54, 1999.

\bibitem{Gnedin:2000uj}
Nickolay~Y. Gnedin.
\newblock {Effect of reionization on the structure formation in the universe}.
\newblock {\em Astrophys.J.}, 542:535--541, 2000.

\bibitem{Dijkstra04}
M.~{Dijkstra}, Z.~{Haiman}, M.~J. {Rees}, and D.~H. {Weinberg}.
\newblock {Photoionization Feedback in Low-Mass Galaxies at High Redshift}.
\newblock {\em \apj}, 601:666--675, February 2004.

\bibitem{Shapiro:2003gxa}
Paul~R. Shapiro, Ilian~T. Iliev, and Alejandro~C. Raga.
\newblock {Photoevaporation of cosmological minihalos during reionization}.
\newblock {\em Mon.Not.Roy.Astron.Soc.}, 348:753, 2004.

\bibitem{Okamoto:2008sn}
Takashi Okamoto, Liang Gao, and Tom Theuns.
\newblock {Massloss of galaxies due to a UV-background}.
\newblock 2008.

\bibitem{Sobacchi:2013wv}
Emanuele Sobacchi and Andrei Mesinger.
\newblock {The depletion of gas in high-redshift dwarf galaxies from an
  inhomogeneous reionization}.
\newblock 2013.

\bibitem{Haiman:1995jy}
Zoltan Haiman, Martin~J. Rees, and Abraham Loeb.
\newblock {H(2) cooling of primordial gas triggered by UV irradiation}.
\newblock {\em Astrophys.J.}, 467:522, 1996.

\bibitem{Pawlik:2008mr}
Andreas~H. Pawlik, Joop Schaye, and Eveline van Scherpenzeel.
\newblock {Keeping the Universe ionised: Photo-ionisation heating and the
  critical star formation rate at redshift z = 6}.
\newblock 2008.

\bibitem{Emberson13}
J.~D. {Emberson}, R.~M. {Thomas}, and M.~A. {Alvarez}.
\newblock {The Opacity of the Intergalactic Medium during Reionization:
  Resolving Small-scale Structure}.
\newblock {\em \apj}, 763:146, February 2013.

\bibitem{McQuinn:2011aa}
Matthew McQuinn, S.~Peng Oh, and Claude-Andre Faucher-Giguere.
\newblock {On Lyman-limit Systems and the Evolution of the Intergalactic
  Ionizing Background}.
\newblock {\em Astrophys.J.}, 743:82, 2011.

\bibitem{2007ApJ...657...15K}
K.~{Kohler}, N.~Y. {Gnedin}, and A.~J.~S. {Hamilton}.
\newblock {Large-Scale Simulations of Reionization}.
\newblock {\em \apj}, 657:15--29, March 2007.

\bibitem{Oh:2003pm}
S.~Peng Oh and Zoltan Haiman.
\newblock {Fossil HII regions: Self-limiting star formation at high redshift}.
\newblock {\em Mon.Not.Roy.Astron.Soc.}, 346:456, 2003.

\bibitem{Furlanetto:2006tf}
Steven Furlanetto.
\newblock {The Global 21 Centimeter Background from High Redshifts}.
\newblock {\em Mon.Not.Roy.Astron.Soc.}, 371:867--878, 2006.

\bibitem{Ade:2013zuv}
P.A.R. Ade et~al.
\newblock {Planck 2013 results. XVI. Cosmological parameters}.
\newblock 2013.

\bibitem{Bennett:2012zja}
C.L. Bennett et~al.
\newblock {Nine-Year Wilkinson Microwave Anisotropy Probe (WMAP) Observations:
  Final Maps and Results}.
\newblock {\em Astrophys.J.Suppl.}, 208:20, 2013.

\bibitem{2010ApJ...709L.133B}
R.~J. {Bouwens}, G.~D. {Illingworth}, P.~A. {Oesch}, M.~{Stiavelli}, P.~{van
  Dokkum}, M.~{Trenti}, D.~{Magee}, I.~{Labb{\'e}}, M.~{Franx}, C.~M.
  {Carollo}, and V.~{Gonzalez}.
\newblock {Discovery of z \~{} 8 Galaxies in the Hubble Ultra Deep Field from
  Ultra-Deep WFC3/IR Observations}.
\newblock {\em \apjl}, 709:L133--L137, February 2010.

\bibitem{2010MNRAS.409..855B}
A.~J. {Bunker}, S.~{Wilkins}, R.~S. {Ellis}, D.~P. {Stark}, S.~{Lorenzoni},
  K.~{Chiu}, M.~{Lacy}, M.~J. {Jarvis}, and S.~{Hickey}.
\newblock {The contribution of high-redshift galaxies to cosmic reionization:
  new results from deep WFC3 imaging of the Hubble Ultra Deep Field}.
\newblock {\em \mnras}, 409:855--866, December 2010.

\bibitem{Robertson:2013bq}
Brant~E. Robertson, Steven~R. Furlanetto, Evan Schneider, Stephane Charlot,
  Richard~S. Ellis, et~al.
\newblock {New Constraints on Cosmic Reionization from the 2012 Hubble Ultra
  Deep Field Campaign}.
\newblock {\em Astrophys.J.}, 768:71, 2013.

\bibitem{2014arXiv1410.5439F}
S.~L. {Finkelstein}, R.~E. {Ryan}, Jr., C.~{Papovich}, M.~{Dickinson},
  M.~{Song}, R.~{Somerville}, H.~C. {Ferguson}, B.~{Salmon}, M.~{Giavalisco},
  A.~M. {Koekemoer}, M.~L.~N. {Ashby}, P.~{Behroozi}, M.~{Castellano}, J.~S.
  {Dunlop}, S.~M. {Faber}, G.~G. {Fazio}, A.~{Fontana}, N.~A. {Grogin},
  N.~{Hathi}, J.~{Jaacks}, D.~D. {Kocevski}, R.~{Livermore}, R.~J. {McLure},
  E.~{Merlin}, B.~{Mobasher}, J.~A. {Newman}, M.~{Rafelski}, V.~{Tilvi}, and
  S.~P. {Willner}.
\newblock {The Evolution of the Galaxy Rest-Frame Ultraviolet Luminosity
  Function Over the First Two Billion Years}.
\newblock {\em ArXiv e-prints}, October 2014.

\bibitem{1976ApJ...203..297S}
P.~{Schechter}.
\newblock {An analytic expression for the luminosity function for galaxies.}
\newblock {\em \apj}, 203:297--306, January 1976.

\bibitem{Bruzual:2003tq}
G.~Bruzual and Stephane Charlot.
\newblock {Stellar population synthesis at the resolution of 2003}.
\newblock {\em Mon.Not.Roy.Astron.Soc.}, 344:1000, 2003.

\bibitem{MiraldaEscude:2002yd}
Jordi Miralda-Escude.
\newblock {On the evolution of the ionizing emissivity of galaxies and quasars
  required by the hydrogen reionization}.
\newblock {\em Astrophys.J.}, 597:66--73, 2003.

\bibitem{2007ApJ...670..928B}
R.~J. {Bouwens}, G.~D. {Illingworth}, M.~{Franx}, and H.~{Ford}.
\newblock {UV Luminosity Functions at z\~{}4, 5, and 6 from the Hubble Ultra
  Deep Field and Other Deep Hubble Space Telescope ACS Fields: Evolution and
  Star Formation History}.
\newblock {\em \apj}, 670:928--958, December 2007.

\bibitem{2013MNRAS.432.2696M}
R.~J. {McLure}, J.~S. {Dunlop}, R.~A.~A. {Bowler}, E.~{Curtis-Lake},
  M.~{Schenker}, R.~S. {Ellis}, B.~E. {Robertson}, A.~M. {Koekemoer}, A.~B.
  {Rogers}, Y.~{Ono}, M.~{Ouchi}, S.~{Charlot}, V.~{Wild}, D.~P. {Stark}, S.~R.
  {Furlanetto}, M.~{Cirasuolo}, and T.~A. {Targett}.
\newblock {A new multifield determination of the galaxy luminosity function at
  z = 7-9 incorporating the 2012 Hubble Ultra-Deep Field imaging}.
\newblock {\em \mnras}, 432:2696--2716, July 2013.

\bibitem{2013ApJ...768..196S}
M.~A. {Schenker}, B.~E. {Robertson}, R.~S. {Ellis}, Y.~{Ono}, R.~J. {McLure},
  J.~S. {Dunlop}, A.~{Koekemoer}, R.~A.~A. {Bowler}, M.~{Ouchi},
  E.~{Curtis-Lake}, A.~B. {Rogers}, E.~{Schneider}, S.~{Charlot}, D.~P.
  {Stark}, S.~R. {Furlanetto}, and M.~{Cirasuolo}.
\newblock {The UV Luminosity Function of Star-forming Galaxies via Dropout
  Selection at Redshifts z \~{} 7 and 8 from the 2012 Ultra Deep Field
  Campaign}.
\newblock {\em \apj}, 768:196, May 2013.

\bibitem{Bolton:2007b}
James~S. Bolton and Martin~G. Haehnelt.
\newblock {The observed ionization rate of the intergalactic medium and the
  ionizing emissivity at z>=5: Evidence for a photon starved and extended epoch
  of reionization}.
\newblock {\em Mon.Not.Roy.Astron.Soc.}, 382:325, 2007.

\bibitem{2010ApJ...721.1448S}
A.~{Songaila} and L.~L. {Cowie}.
\newblock {The Evolution of Lyman Limit Absorption Systems to Redshift Six}.
\newblock {\em \apj}, 721:1448--1466, October 2010.

\bibitem{Worseck:2014fya}
G‡bor Worseck, J.~Xavier Prochaska, John~M. O'Meara, George~D. Becker, Sara
  Ellison, et~al.
\newblock {The Giant Gemini GMOS survey of z>4.4 quasars - I. Measuring the
  mean free path across cosmic time}.
\newblock {\em Mon.Not.Roy.Astron.Soc.}, 445:1745, 2014.

\bibitem{Kuhlen:2012vy}
M.~Kuhlen and C.A. Faucher-Giguere.
\newblock {Concordance models of reionization: implications for faint galaxies
  and escape fraction evolution}.
\newblock 2012.

\bibitem{Becker:2013ffa}
George~D. Becker and James~S. Bolton.
\newblock {New Measurements of the Ionizing Ultraviolet Background over 2 < z <
  5 and Implications for Hydrogen Reionization}.
\newblock 2013.

\bibitem{Alvarez12}
M.~A. {Alvarez}, K.~{Finlator}, and M.~{Trenti}.
\newblock {Constraints on the Ionizing Efficiency of the First Galaxies}.
\newblock {\em \apjl}, 759:L38, November 2012.

\bibitem{Choudhury:2008aw}
T.~Roy Choudhury, M.G. Haehnelt, and J.~Regan.
\newblock {Inside-out or Outside-in: The topology of reionization in the
  photon-starved regime suggested by Lyman-alpha forest data}.
\newblock 2008.

\bibitem{Mesinger12}
A.~{Mesinger}, M.~{McQuinn}, and D.~N. {Spergel}.
\newblock {The kinetic Sunyaev-Zel'dovich signal from inhomogeneous
  reionization: a parameter space study}.
\newblock {\em \mnras}, 422:1403--1417, May 2012.

\bibitem{Sobacchi:2014rua}
Emanuele Sobacchi and Andrei Mesinger.
\newblock {Inhomogeneous recombinations during cosmic reionization}.
\newblock 2014.

\bibitem{MiraldaEscude:1998qs}
Jordi Miralda-Escude, Martin Haehnelt, and Martin~J. Rees.
\newblock {Reionization of the inhomogeneous universe}.
\newblock {\em Astrophys.J.}, 530:1--16, 2000.

\bibitem{Furlanetto:2005xx}
Steven~R. Furlanetto and S.~Peng Oh.
\newblock {Taxing the rich: Recombinations and bubble growth during
  reionization}.
\newblock {\em Mon.Not.Roy.Astron.Soc.}, 363:1031--1048, 2005.

\bibitem{Schaye:2001me}
Joop Schaye.
\newblock {Model independent insights into the nature of the Lyman-alpha forest
  and the distribution of matter in the universe}.
\newblock {\em Astrophys.J.}, 559:507, 2001.

\bibitem{2011ApJ...737L..37A}
G.~{Altay}, T.~{Theuns}, J.~{Schaye}, N.~H.~M. {Crighton}, and C.~{Dalla
  Vecchia}.
\newblock {Through Thick and Thin - H I Absorption in Cosmological
  Simulations}.
\newblock {\em \apjl}, 737:L37, August 2011.

\bibitem{2013MNRAS.430.2427R}
A.~{Rahmati}, A.~H. {Pawlik}, M.~{Raicevic}, and J.~{Schaye}.
\newblock {On the evolution of the H I column density distribution in
  cosmological simulations}.
\newblock {\em \mnras}, 430:2427--2445, April 2013.

\bibitem{Haardt:2001zf}
Francesco Haardt, Piero Madaus, and Piero Madau.
\newblock {Modeling the uv/x-ray cosmic background with cuba}.
\newblock 2001.

\bibitem{Furlanetto:2004ha}
Steven Furlanetto, Matias Zaldarriaga, and Lars Hernquist.
\newblock {Statistical probes of reionization with 21 cm tomography}.
\newblock {\em Astrophys.J.}, 613:16--22, 2004.

\bibitem{Mellema:2006pd}
Garrelt Mellema, Ilian~T. Iliev, Ue-Li Pen, and Paul~R. Shapiro.
\newblock {Simulating cosmic reionization at large scales. 2. the 21-cm
  emission features and statistical signals}.
\newblock {\em Mon.Not.Roy.Astron.Soc.}, 372:679--692, 2006.

\bibitem{Lidz:2007az}
Adam Lidz, Oliver Zahn, Matthew McQuinn, Matias Zaldarriaga, and Lars
  Hernquist.
\newblock {Detecting the Rise and Fall of 21 cm Fluctuations with the Murchison
  Widefield Array}.
\newblock {\em Astrophys.J.}, 680:962--974, 2008.

\bibitem{McQuinn:2007dy}
Matthew McQuinn, Lars Hernquist, Matias Zaldarriaga, and Suvendra Dutta.
\newblock {Studying Reionization with Ly-alpha Emitters}.
\newblock {\em Mon.Not.Roy.Astron.Soc.}, 381:75--96, 2007.

\bibitem{Lidz:2007mz}
Adam Lidz, Matthew McQuinn, and Matias Zaldarriaga.
\newblock {Quasar Proximity Zones and Patchy Reionization}.
\newblock {\em Astrophys.J.}, 670:39--59, 2007.

\bibitem{Mesinger:2009mv}
Andrei Mesinger.
\newblock {Was reionization complete by z ~ 5-6?}
\newblock 2009.

\bibitem{Malloy:2014tba}
Matthew Malloy and Adam Lidz.
\newblock {How to Search for Islands of Neutral Hydrogen in the $z \sim 5.5$
  IGM}.
\newblock 2014.

\bibitem{McQuinn:2007gm}
Matthew McQuinn, Adam Lidz, Matias Zaldarriaga, Lars Hernquist, and Suvendra
  Dutta.
\newblock {Probing the Neutral Fraction of the IGM with GRBs during the Epoch
  of Reionization}.
\newblock {\em Mon.Not.Roy.Astron.Soc.}, 388:1101--1110, 2008.

\bibitem{Mesinger:2007kd}
Andrei Mesinger and Steven Furlanetto.
\newblock {Lyman-alpha Damping Wing Constraints on Inhomogeneous Reionization}.
\newblock {\em Mon.Not.Roy.Astron.Soc.}, 385:1348, 2008.

\bibitem{Barkana:2003qk}
Rennan Barkana and Abraham Loeb.
\newblock {Unusually large fluctuations in the statistics of galaxy formation
  at high redshift}.
\newblock {\em Astrophys.J.}, 609:474--481, 2004.

\bibitem{Mesinger:2007pd}
Andrei Mesinger and Steven Furlanetto.
\newblock {Efficient Simulations of Early Structure Formation and
  Reionization}.
\newblock {\em Astrophys.J.}, 669:663, 2007.

\bibitem{McQuinn:2006et}
Matthew McQuinn, Adam Lidz, Oliver Zahn, Suvendra Dutta, Lars Hernquist, et~al.
\newblock {The Morphology of HII Regions during Reionization}.
\newblock {\em Mon.Not.Roy.Astron.Soc.}, 377:1043--1063, 2007.

\bibitem{2011MNRAS.414..727Z}
O.~{Zahn}, A.~{Mesinger}, M.~{McQuinn}, H.~{Trac}, R.~{Cen}, and L.~E.
  {Hernquist}.
\newblock {Comparison of reionization models: radiative transfer simulations
  and approximate, seminumeric models}.
\newblock {\em \mnras}, 414:727--738, June 2011.

\bibitem{Trac:2006vr}
Hy~Trac and Renyue Cen.
\newblock {Radiative transfer simulations of cosmic reionization. 1.
  Methodology and initial results}.
\newblock {\em Astrophys.J.}, 2006.

\bibitem{Lidz:2006vj}
Adam Lidz, Oliver Zahn, Matthew McQuinn, Matias Zaldarriaga, and Suvendra
  Dutta.
\newblock {Higher Order Contributions to the 21 cm Power Spectrum}.
\newblock {\em Astrophys.J.}, 659:865--876, 2007.

\bibitem{1965ApJ...142.1633G}
J.~E. {Gunn} and B.~A. {Peterson}.
\newblock {On the Density of Neutral Hydrogen in Intergalactic Space.}
\newblock {\em \apj}, 142:1633--1641, November 1965.

\bibitem{Meiksin:2003qb}
Avery Meiksin and 1~White, Martin~J.
\newblock {The Effects of UV background correlations on Ly-alpha forest flux
  statistics}.
\newblock {\em Mon.Not.Roy.Astron.Soc.}, 350:1107, 2004.

\bibitem{Croft:2003qn}
Rupert~A.C. Croft.
\newblock {Ionizing radiation fluctuations and large scale structure in the
  Lyman-alpha forest}.
\newblock {\em Astrophys.J.}, 610:642--662, 2004.

\bibitem{McDonald:2004xp}
Patrick McDonald, Uros Seljak, Renyu Cen, Paul Bode, and Jeremiah~P. Ostriker.
\newblock {Physical effects on the Ly-alpha forest flux power spectrum: Damping
  wings, ionizing radiation fluctuations, and galactic winds}.
\newblock {\em Mon.Not.Roy.Astron.Soc.}, 360:1471--1482, 2005.

\bibitem{Zuo92}
L.~{Zuo}.
\newblock {Fluctuations in the ionizing background}.
\newblock {\em \mnras}, 258:36--44, September 1992.

\bibitem{Zuo93}
L.~{Zuo} and E.~S. {Phinney}.
\newblock {Absorption by Discrete Intergalactic Clouds: Theory and Some
  Applications}.
\newblock {\em \apj}, 418:28, November 1993.

\bibitem{2009MNRAS.400.1461M}
A.~{Mesinger} and S.~{Furlanetto}.
\newblock {The inhomogeneous ionizing background following reionization}.
\newblock {\em \mnras}, 400:1461--1471, December 2009.

\bibitem{Crociani11}
D.~{Crociani}, A.~{Mesinger}, L.~{Moscardini}, and S.~{Furlanetto}.
\newblock {The distribution of Lyman-limit absorption systems during and after
  reionization}.
\newblock {\em \mnras}, 411:289--300, February 2011.

\bibitem{Cooray:2002dia}
Asantha Cooray and Ravi~K. Sheth.
\newblock {Halo models of large scale structure}.
\newblock {\em Phys.Rept.}, 372:1--129, 2002.

\bibitem{Becker:2014oga}
George~D. Becker, James~S. Bolton, Piero Madau, Max Pettini, Emma~V.
  Ryan-Weber, et~al.
\newblock {Evidence of patchy hydrogen reionization from an extreme Ly$\alpha$
  trough below redshift six}.
\newblock 2014.

\bibitem{Lidz:2005gn}
Adam Lidz, S.~Peng Oh, and Steven~R. Furlanetto.
\newblock {Have we detected patchy reionization in quasar spectra?}
\newblock {\em Astrophys.J.}, 639:L47--L60, 2006.

\bibitem{Theuns:2002yc}
Tom Theuns, Joop Schaye, Saleem Zaroubi, Tae-Sun Kim, Panayiotis Tzanavaris,
  et~al.
\newblock {Constraints on reionization from the thermal history of the
  intergalactic medium}.
\newblock {\em Astrophys.J.}, 567:L103, 2002.

\bibitem{Hui:2003hn}
Lam Hui and Zoltan Haiman.
\newblock {The Thermal memory of reionization history}.
\newblock {\em Astrophys.J.}, 596:9--18, 2003.

\bibitem{Abel:1999vj}
Tom Abel and Martin~G. Haehnelt.
\newblock {Radiative transfer effects during photoheating of the intergalactic
  medium}.
\newblock {\em Astrophys.J.}, 520:L13--L16, 1999.

\bibitem{Furlanetto:2009kr}
Steven Furlanetto and S.~Peng Oh.
\newblock {The Equation of State of the Intergalactic Medium After Hydrogen
  Reionization}.
\newblock 2009.

\bibitem{Lidz:2014jxa}
Adam Lidz and Matthew Malloy.
\newblock {On Modeling and Measuring the Temperature of the z~5 intergalactic
  medium}.
\newblock {\em Astrophys.J.}, 788:175, 2014.

\bibitem{1970A&A.....5...84Z}
Y.~B. {Zel'dovich}.
\newblock {Gravitational instability: An approximate theory for large density
  perturbations.}
\newblock {\em \aap}, 5:84--89, March 1970.

\bibitem{Trac:2008yz}
Hy~Trac, Renyue Cen, and Abraham Loeb.
\newblock {Imprint of Inhomogeneous Hydrogen Reionization on the Temperature
  Distribution of the Intergalactic Medium}.
\newblock {\em Astrophys.J.}, 689:L81--L84, 2008.

\bibitem{Becker11}
G.~D. {Becker}, J.~S. {Bolton}, M.~G. {Haehnelt}, and W.~L.~W. {Sargent}.
\newblock {Detection of extended He II reionization in the temperature
  evolution of the intergalactic medium}.
\newblock {\em \mnras}, 410:1096--1112, January 2011.

\bibitem{Oh:2004rm}
S.~Peng Oh and Steven~R. Furlanetto.
\newblock {How universal is the Gunn-Peterson trough at z ~ 6? A Closer look at
  the quasar SDSS J1148+5251}.
\newblock {\em Astrophys.J.}, 620:L9--L12, 2005.

\bibitem{2009MNRAS.398L..26B}
J.~S. {Bolton} and G.~D. {Becker}.
\newblock {Resolving the high redshift Ly{$\alpha$} forest in smoothed particle
  hydrodynamics simulations}.
\newblock {\em \mnras}, 398:L26--L30, September 2009.

\end{thebibliography}

\end{document}